\documentclass[twocolumn,showpacs,amsmath,amssymb,prb,footinbib,floatfix,superscriptaddress]{revtex4}
\usepackage{graphicx}
\usepackage{savesym}
\usepackage{amsfonts}
\usepackage{bm}
\usepackage{color}
\usepackage{hyperref}
\usepackage{subfigure}
\usepackage{wasysym}
\usepackage{upgreek}
\usepackage{float}

\begin{document}

\title{Tripartite entangled plaquette state in a cluster magnet}

\author{Juan Carrasquilla}
\email{jcarrasquilla@dwavesys.com} 
\affiliation{D-Wave Systems Inc., 3033 Beta Avenue, Burnaby BC Canada V5G 4M9}
\affiliation{Perimeter Institute for Theoretical Physics, Waterloo, Ontario, N2L 2Y5, Canada} 

\author{Gang Chen}
\affiliation{State Key Laboratory of Surface Physics, Center for Field Theory and Particle Physics, 
Department of Physics, Collaborative Innovation Center of Advanced Microstructures,
Fudan University, Shanghai 200433, People's Republic of China}
\affiliation{Perimeter Institute for Theoretical Physics, Waterloo, Ontario, N2L 2Y5, Canada}

\author{Roger G. Melko} 
\affiliation{Perimeter Institute for Theoretical Physics, Waterloo, Ontario, N2L 2Y5, Canada} 
\affiliation{Department of Physics and Astronomy, University of Waterloo, Ontario, N2L 3G1, Canada}

\date{\today}

\begin{abstract}
{
Using large-scale quantum Monte Carlo simulations we show 
that a spin-$1/2$ XXZ model on a two-dimensional anisotropic 
Kagome lattice exhibits a tripartite entangled plaquette state 
that preserves all of the Hamiltonian symmetries.  It is connected
via phase boundaries to a ferromagnet and a valence-bond solid that break
U(1) and lattice translation symmetries, respectively. We study the phase diagram of the 
model in detail, in particular the transitions to the tripartite entangled plaquette 
state, which are consistent with conventional order-disorder transitions. 
Our results can be interpreted as a description of the charge sector 
dynamics of a Hubbard model applied to the description of the spin 
liquid candidate ${\mathrm{LiZn}}_{2}{\mathrm{Mo}}_{3}{\mathrm{O}}_{8}$, 
as well as a model of strongly correlated bosonic 
atoms loaded onto highly tunable {\it trimerized} optical 
Kagome lattices.
}
\end{abstract}

\maketitle

\section{\uppercase{Introduction}} %

Frustration in the context of magnetism refers to 
the phenomena where interactions between magnetic moments 
compete at the microscopic level, usually due to 
a combination of antiferromagnetic exchange and 
lattice geometry. Unlike unfrustrated systems, 
where symmetry breaking in the groundstate prevails, 
the inability of frustrated magnets to satisfy 
each and every microscopic interaction can lead to the 
emergence of exotic groundstate phases, such as
valence bond solids,\cite{Tamura2006,Matan2010,Sachdev2011,Vasiliev2015} 
spin liquids,\cite{Pratt2011,Han2012,Balents2010,Yan2011}
classical\cite{Bramwell16112001} and quantum spin 
ices,\cite{Gingras_McClarty2013,Ross2011} among many others. 

Geometric frustration typically arises in magnetic moments 
that are localized on single ions, through their spatial arrangement
and that of their exchange interactions.  However, moments on 
ionic crystal lattices are often susceptible to
structural distortions, orbital couplings, or mixing between 
magnetic and non-magnetic layers.  These and other perturbations
may disrupt the formation of delicate exotic phases, 
leading in many cases to conventional ordering.
This difficulty may be alleviated in the recently-discovered 
geometrically frustrated antiferromagnets where the magnetic 
moments are localized on small transition-metal clusters, 
rather than being localized on a single ion.\cite{Sheckelton2012}
These materials have been shown to avoid the key limitations 
of ion-localized moments mentioned above, 
making them natural candidates to search for exotic states of matter.\cite{Sheckelton2012,Mourigal2014,Sheckelton2014,Haraguchi2015,Chen2014,Chen2015}
In a recent development, Sheckelton {\it et al}\cite{Sheckelton2012,Mourigal2014,Sheckelton2014} 
found that the molecular magnet on the triangular lattice 
${\mathrm{LiZn}}_{2}{\mathrm{Mo}}_{3}{\mathrm{O}}_{8}$ exhibits spin liquid 
behavior with low-energy spin correlations consistent with 
the highly-coveted resonating valence-bond solid state.\cite{ANDERSON1987} 
Similarly, evidence of strong quantum fluctuations and 
spin liquid behavior was found in a related cluster magnet 
${\mathrm{Li}}_{2}{\mathrm{ScMo}}_{3}{\mathrm{O}}_{8}$, 
while the isomorphic compound ${\mathrm{Li}}_{2}{\mathrm{InMo}}_{3}{\mathrm{O}}_{8}$ 
was found to develop long-range 120$^{\circ}$ magnetic order.\cite{Haraguchi2015} 

Attempts to elucidate the microscopic origin of the experimental 
observations in ${\mathrm{LiZn}}_{2}{\mathrm{Mo}}_{3}{\mathrm{O}}_{8}$ have
included a model of lattice distortions leading to an emergent 
honeycomb lattice where the spins form a quantum 
spin liquid,\cite{Flint2013} as well as a purely electronic 
description based on a $1/6$-filled extended Hubbard model with 
nearest-neighbor repulsion on a trimerized Kagome lattice.\cite{Chen2015} 
The later work suggested that the ground state 
of ${\mathrm{LiZn}}_{2}{\mathrm{Mo}}_{3}{\mathrm{O}}_{8}$ may 
be a U(1) spin liquid with plaquette charge order and 
a spinon Fermi surface, whose finite-temperature properties 
may explain the two surprising Curie-Weiss regimes observed in the 
experimental data.\cite{Chen2015} 

The U(1) spin liquid state arises from a generic procedure 
where a mean-field decoupling of the charge and spin 
degrees of freedom in terms of a slave-rotor representation
of the electron operators\cite{Florens2004} is performed.  
In such an approach, the electron systems are mapped onto a 
spinon Hamiltonian coupled to a bosonic lattice model of the charge
sector via mean-field parameters. 
The intuition behind such an approach is in the observation
that in certain strongly-coupled electron systems, the dynamics 
of the spin and charge degrees of freedom is markedly different. 
Therefore, the electron may be better understood as being composed of 
separate charge and spin variables. To determine the fate the ground 
state of the overall fermionic system, both spinon and bosonic Hamiltonians 
have to be solved simultaneously. Since the resulting bosonic Hamiltonian 
associated with the charge sector is generically strongly interacting, such a  
problem can be solved, for instance, via a standard mean-field 
decoupling. However, other approaches that include some spatial 
correlations, e.g., quantum Monte Carlo (QMC), exact diagonalization,  
or density-matrix renormalization group methods, are desirable and clearly 
improve the quality of the description of the many-body 
electron problem under study.\cite{Zhao2007}

\begin{figure*}[t]
\includegraphics[trim=0 0 0 0, clip,width=0.8\textwidth]{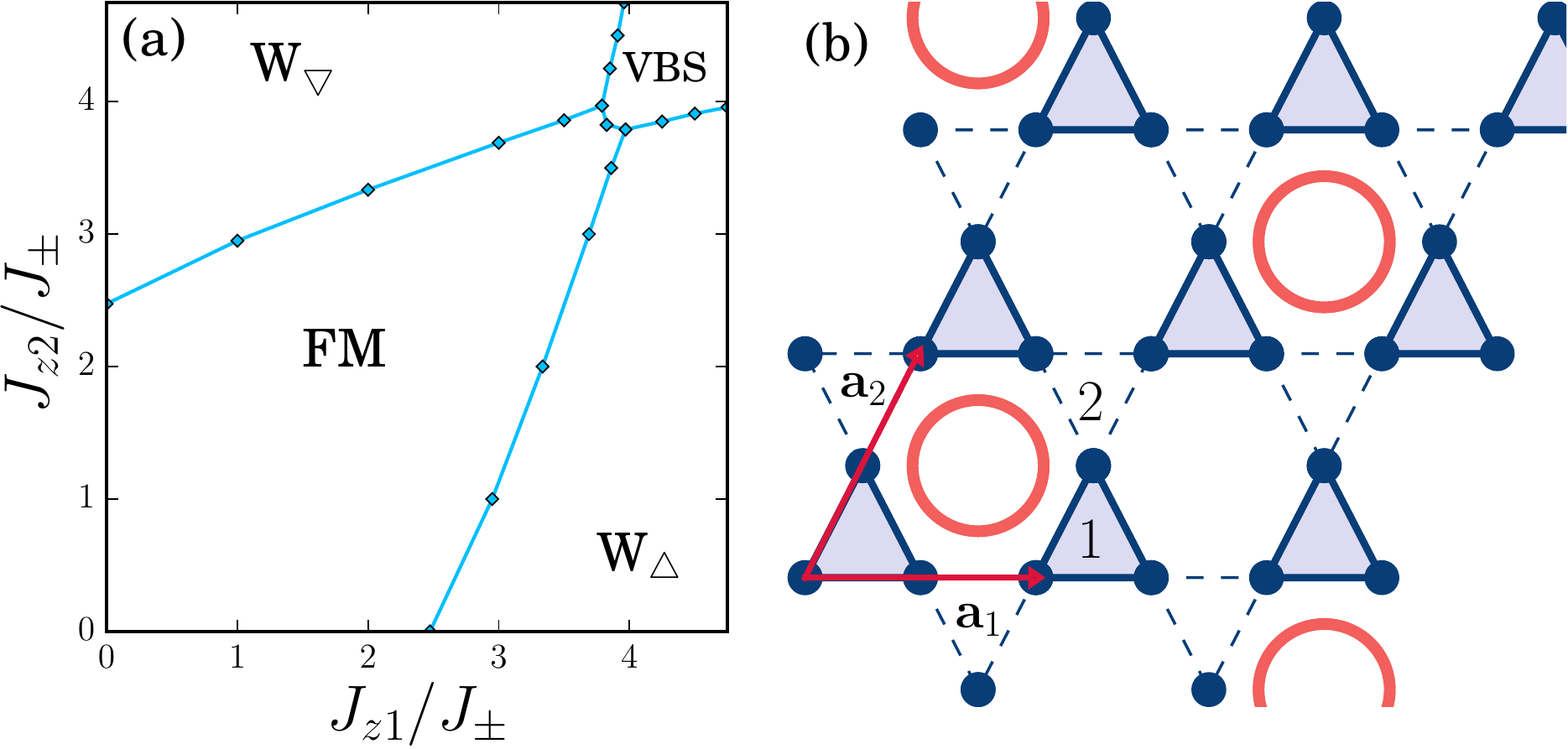}
\caption{(Color online.) 
(a) Ground state phase diagram of the Hamiltonian 
in Eq.(\ref{HAM}) on the trimerized Kagome lattice as a 
function of $J_{z1}/J_{\pm}$ and $J_{z2}/J_{\pm}$
at fixed magnetization $m=-1/6$. 
(b) Our simulations are defined on periodic tori 
of size $V=N_{\text{s}}\times3$ spanned by the primitive 
vectors $\mathbf{a}_1$ and $\mathbf{a}_2$, 
where $\lVert\mathbf{a}_1\rVert=\lVert\mathbf{a}_2\rVert=2$. 
$N_{\text{s}}$ is the number of sites of the underlying triangular 
lattice of the Kagome lattice. The system contains three phases: 
a ferromagnet (FM), a valence-bond solid (VBS), and a tripartite entangled plaquette state (W). 
In the hardcore-boson language, the FM phase corresponds
to a superfluid (SF), while the W phases correspond to fractionally filled Mott insulators. 
The red circles illustrate the subset of 
hexagons where spins resonantly flip in the VBS phase. 
The index $t=1,2$ indicates the different
up (1, shaded with solid lines) and down (2, white with dashed 
lines) triangles forming the Kagome lattice.}\label{fig:lattice}
\end{figure*}

In this work, we use large-scale QMC simulations 
to show that a two-dimensional spin-$1/2$ model 
on an anisotropic Kagome lattice--proposed as the
description of the charge sector of an extended 
Hubbard model applied to the cluster magnet
${\mathrm{LiZn}}_{2}{\mathrm{Mo}}_{3}{\mathrm{O}}_{8}$--exhibits 
three different phases: a ferromagnet (FM), a fractionally-filled tripartite 
entangled plaquette state (W) and a valence-bond solid (VBS). 
These phases are arranged in the phase diagram presented in Fig.\ref{fig:lattice}(a). 
We find that the FM-to-W insulator transition is 
continuous and belongs to the three-dimensional $O(2)$ 
universality class if the magnetization is kept constant
across the transition;\cite{Fisher1989} 
instead, if the magnetization varies, 
the transition is generically mean-field.\cite{Fisher1989} We also find 
evidence indicating that the transition between the  W state and the VBS, up to 
the system sizes accessible with QMC, appears to be first order, in 
agreement with the theoretical expectation that it belongs 
to the three-dimensional three-state clock universality.\cite{Herrmann1979} 
Finally, we re-examine the transition between the FM and the VBS in terms of a
recently proposed scenario of quantum criticality with two length scales~\cite{Shao213}.
We find that the presence of irregular system-size dependent oscillations in the observables prevents 
us from drawing a firm conclusion about the applicability of this scenario to
the FM-to-VBS transition.

\section{\uppercase{Model and Quantum Monte Carlo simulations}} 
\label{sec2}

We begin by motivating a Hamiltonian for the charge sector of the
cluster magnet ${\mathrm{LiZn}}_{2}{\mathrm{Mo}}_{3}{\mathrm{O}}_{8}$.
This material is formed by small triangular 
${\mathrm{Mo}}_{3}{\mathrm{O}}_{13}$ units, 
where each triangular plaquette accommodates one unpaired electron. 
The small ${\mathrm{Mo}}_{3}{\mathrm{O}}_{13}$ 
triangular units are located on the sites of a triangular 
lattice shown in Fig.~\ref{fig:lattice}(b). 
Thus the ${\mathrm{Mo}}$ atoms can be thought 
of as forming a trimerized Kagome lattice. 
Following Ref.~\onlinecite{Chen2015}, we consider 
the electrons hopping on the Kagome lattice as described by 
an extended Hubbard model at 1/6 electron filling. 
In this model, both the {\it on-site} 
and the {\it nearest-neighbor} Coulomb interactions are included in addition
to the electron hopping. The authors of Ref.~\onlinecite{Chen2015} 
employ a standard slave-rotor representation of the constituent 
fermions.\cite{Florens2004} The electron operator is
reformulated as the product of a U(1) charge rotor variable and
a fermionic spinon, $c^{\dag}_{{\bf r}\sigma}= e^{i\theta_{\bf r}} f^{\dag}_{{\bf r}\sigma}$, 
where the bosonic rotor $e^{i\theta_{\bf r}}$ (the fermionic spinon $f^{\dag}_{{\bf r}\sigma}$) 
creates an electron charge (a spinon with spin $\sigma$) 
at lattice site ${\bf r}$. Since the Hilbert space has been enlarged, 
one has to introduce a constraint to get back to 
the physical Hilbert space through an angular momentum variable: 
$S^{z}_{\bf r}=\sum_{\sigma} f^{\dag}_{{\bf r}\sigma}f^{}_{{\bf r}\sigma} - 1/2$. 
The local electron Hilbert space is thus represented 
as $\lvert 0  \rangle_c=  
{\lvert 0  \rangle}_f {\lvert { S^z =-1/2 } \rangle}_{\theta} $, 
${\lvert \uparrow  \rangle}_c=  
{\lvert \uparrow  \rangle}_f {\lvert S^z =1/2 \rangle}_{\theta}$, 
${\lvert \downarrow  \rangle}_c=  
{\lvert \downarrow  \rangle}_f {\lvert  S^z =1/2 \rangle}_{\theta}$,
and 
${\lvert \uparrow \downarrow \rangle}_c
=  {\lvert \uparrow \downarrow \rangle}_f 
{\lvert S^z =3/2 \rangle}_{\theta}$.  
When the on-site Coulomb interaction is dominant, 
the double electron occupancy on a single site is forbidden and the operator 
$S^z$ describes an effective spin-1/2 
angular momentum operator that is conjugate to the charge rotor 
variable, i.e., $[\theta_{\bf r},S^{z}_{{\bf r}'}]=i\delta_{{\bf r},{\bf r}'}$. 
Thus the rotor operators can be identified as the spin ladder
operators $S^{\pm}_{\bf r}=e^{\pm i \theta_{\bf r}}$. Using a mean-field 
decoupling, the original Hubbard model is transformed into 
two Hamiltonians for the spinon and charge sectors coupled 
via mean-field parameters. 
Under the slave-rotor reformulation, 
the charge sector of the extended Hubbard model is now described by an effective 
spin-1/2 model on an anisotropic Kagome lattice with 
\begin{eqnarray}
\mathcal{H}_{\text{c}} &=& \sum\limits_{\langle  
\text{\bf r}  \text{\bf r}' \rangle, t=1,2 } 
\Big[ {J}_{z t} S^{z}_{\text{\bf r}} S^{z}_{\text{\bf r}'}
- { \frac{J_{\pm t}}{2}}( S^{+}_{\text{\bf r}}   
S^{-}_{\text{\bf r}'} + \text{h.c.})\Big]  \nonumber \\
&-&h_{\text{eff}}\sum\limits_{\text{\bf r} }S^{z}_{\text{\bf r}},
\label{HAM}
\end{eqnarray}
where the Ising exchange interaction (transverse exchange interaction)
accounts for the nearest-neighbor Coulomb interaction (electron tunneling).
The index $t=1,2$ indicates the different up (1, shaded with solid lines) 
and down (2, white with dashed lines) triangles forming the Kagome lattice, 
as shown in Fig.\ref{fig:lattice}(b). 
The anisotropy of the lattice is encoded in the coupling constants 
$J_{\pm t}$, $J_{z t}$, whereas the effective magnetic field 
$h_{\text{eff}}$ controls the average effective magnetization. 
We will consider systems at both strictly conserved magnetization 
as well as at fixed magnetic fields. 
We set $J_{\pm 1}=J_{\pm 2}=J_{\pm}$ as the reference energy scale of the problem. 
This model can also be thought of as one describing hardcore 
bosons\cite{Matsuda01011957} loaded on a trimerized Kagome lattice. 
In the hardcore-boson language, the FM phase corresponds to a superfluid (SF), 
while the W phases correspond to fractionally-filled Mott insulators where 
a hardcore bosons are localized on the triangles with the largest $J_z$. Such 
a system of hardcore bosons on trimerized lattice geometry can be realized using 
superlattice techniques in ultracold gases.\cite{Santos2004}

The Hamiltonian in Eq.(\ref{HAM}) cannot be solved  
analytically, but large-scale QMC simulations are allowed 
in the sign-problem free regime where $J_{\pm t}>0$. 
This is the natural choice if we interpret Eq.(\ref{HAM}) 
as a system of hardcore bosons. 
We develop a finite-temperature Stochastic Series 
Expansion~\cite{sandvik1999,Syljuaasen2002,Melko2007} (SSE) 
QMC algorithm with {\it directed} loop updates. 
We map out the phase diagram of the model through 
measurements such as magnetization 
$m_z=\langle \hat{m}\rangle=\langle\frac{1}{V}
\sum_{\text{\bf r}}S_{\text{\bf r}}^{z}\rangle$, 
uniform spin susceptibility 
$\chi_z=\frac{V}{T}\left(\langle \hat{m}^{2} \rangle 
-\langle \hat{m} \rangle^2 \right)$, 
superfluid stiffness $\rho_s$,\cite{Pollock1987} 
and a set of diagonal\cite{sandvik1999} 
and off-diagonal\cite{Dorneich2001} spin structure factors. 
Figure \ref{fig:lattice}(a) shows the QMC phase diagram 
for the model of Eq.~(\ref{HAM}) extracted from 
the finite-temperature and the finite-size scaling of 
the superfluid stiffness and the diagonal structure factor, 
performed up to lattice sizes of $ V = L \times L \times 3 
= 60 \times 60 \times 3$ and inverse temperature of up to 
$\beta = J_{\pm}/T = 60$. The phase diagram 
presented in Fig.\ref{fig:lattice}(a) is obtained at 
strictly enforced fixed magnetization $m_z=-1/6$ that 
corresponds to the 1/6 electron filling in LiZn$_2$Mo$_3$O$_8$.
To do this efficiently, we first tune the magnetic field $h_{\text{eff}}$ 
such that the average magnetization is as close as possible 
to the desired magnetization sector. We subsequently run 
simulations that are still grand-canonical 
but whose measurements are taken only at  
configurations that are in the desired magnetization sector. 

The grand-canonical phase diagram of the isotropic 
$J_{z1}=J_{z2}$ case has been explored in 
Ref.~\onlinecite{Isakov2006} where the authors found two phases:  
a FM and a VBS phase. It was found that the FM phase is 
characterized by long-range in-plane magnetic order 
with wave vector $\text{\bf q}={\bf 0}$, finite superfluid 
stiffness $\rho_s>0$, and uniform susceptibility $\chi_z>0$. 
The VBS was studied in great detail in Ref.~\onlinecite{Isakov2006}: 
it is a gapped, translationally broken 
three-fold degenerate\cite{Nikolic2005} phase, where spins 
resonantly flip in each hexagon marked by red circles in 
Fig.(\ref{fig:lattice})(b). The remaining spins 
anti-align along the $z$ direction and their wave vector 
is $\text{\bf q}=\text{\bf K}=(2\pi/3,0)$.\cite{Isakov2006} 
The average magnetization is $m_z=-1/6$ and $\chi_z=0$.

\section{The tripartite entangled plaquette state }
\label{Winsul}

The gapped, tripartite entangled plaquette states W preserve all the symmetries of the Hamiltonian
and are not present in the isotropic $J_{z1}=J_{z2}$ case\cite{Isakov2006}. 
We find that, in the strong triangle limit, the W states reduce to simple 
product states where the three spins on the strong 
triangle (up $\text{W}_\bigtriangleup$ if $J_{z1}\gg J_{z2}$ 
or down $\text{W}_\bigtriangledown$ if $J_{z2}\gg J_{z1}$) 
form tripartite entangled three-qubit W states.\cite{Santos2004,Damski2005,Durg2000} 
The three-qubit states are given by 
$|\text{W}\rangle=\left( 
|{\uparrow \downarrow\downarrow }\rangle + 
|{\downarrow \uparrow \downarrow}\rangle +
|{\downarrow\downarrow\uparrow } \rangle  
\right)/\sqrt{3}$.

To understand the properties of the W phases and
the phase transitions to the nearby FM and VBS phases,  
we measure several correlation functions. 
We measure the diagonal spin structure factor  
$S^{\alpha\beta}_{\text{\bf q}}/N_{\text{s}} 
=\langle S^{\alpha}_{\text{\bf q}}  
S^{\beta}_{-\text{\bf q}} \rangle -\langle 
S^{\alpha}_{\text{\bf q}}\rangle \langle 
S^{\beta}_{-\text{\bf q}} \rangle$, 
where 
\begin{equation}
S^{\alpha}_{\text{\bf q}}=\frac{1}{N_{\text{s}}} 
\sum\limits_{\text{\bf r}_i} e^{ i\text{\bf q} \cdot
\left(\text{\bf r}_i
+ \boldsymbol{\upalpha} \right) } 
S^{z}_{\text{\bf r}_i+ \boldsymbol{\upalpha}},
\end{equation}
and $N_s$ is the total number of unit cells.
Here, $\text{\bf r}_i$ points to the sites of 
the underlying triangular Bravais lattice and the vector
$\boldsymbol{\upalpha}$ refers to the position of each  
site within the unit cell with respect to the vector 
$\text{\bf r}_i$. Its purpose is to detect the long-range 
diagonal order: if the system magnetically orders then 
$S_{\text{\bf q}}=\sum_{\alpha}S^{\alpha\alpha}_{\text{\bf q}}$ 
will scale with system size for at least one value of $\text{\bf q}$.  
We also consider the off-diagonal spin structure factor~\cite{Dorneich2001}
\begin{equation}
\label{momdist}
n^{\alpha\beta}_{\text{\bf q}} = \frac{1}{N_{\text{s}}}
\sum\limits_{\text{\bf r}_i  \text{\bf r}_j}  
e^{i \text{\bf q} \cdot \left[ \left(  \text{\bf r}_i+ 
\boldsymbol{\upalpha} \right) 
-   \left(  \text{\bf r}_j+ \boldsymbol{\upbeta}  \right)   \right]} 
\langle S^{+}_{ \text{\bf r}_i+ \boldsymbol{\upalpha} } 
S^{-}_{ \text{\bf r}_j+ \boldsymbol{\upbeta}}\rangle.
\end{equation}
Using this quantity, we also study the equivalent of the 
condensate fraction in Bose systems,\cite{Penrose1956,Giamarchi2008} 
defined as the ratio of the ``zero-momentum occupation'' to the volume of the 
system $f_{0}=n_{\text{\bf 0}}/V=\sum\limits_{\alpha} n^{\alpha,\alpha}_{\text{\bf 0}}/V$. 
Finally, we consider the bond-bond structure factor using a 
four-point correlation function
\begin{equation}
BB^{\alpha\beta}_{\text{\bf q}}=
\frac{1}{N_{\text{s}}}\sum_{\text{\bf r}_a\text{\bf r}_b} e^{i\text{\bf q}\left( \text{\bf r}_a -\text{\bf r}_b \right) }\langle B^{\alpha}_{\text{\bf r}_a} B^{\beta}_{\text{\bf r}_b} \rangle,
\end{equation}
where $B^{\alpha}_{\text{\bf r}_a}=S_{i_{a\alpha}}^{+}S_{j_{a\alpha}}^{-}+S_{i_{a\alpha}}^{-}S_{j_{a\alpha}}^{+}$. Nearest neighbor sites $i_{a\alpha}$ and 
$j_{a\alpha}$ belong to one of 6 bonds $\alpha$ in a unit cell 
located at position $\text{\bf r}_a$. If bond order develops then 
$BB_{\text{\bf q}}=\sum_{\alpha}BB^{\alpha \alpha}_{\text{\bf q}}$ 
should scale with system size for at least one value of $\text{\bf q}$, 
with which we define $B_{\text{\bf q}}=BB_{\text{\bf q}}/V$.

\begin{figure*}[t]
\includegraphics[trim=0 0 0 0, clip,width=0.8\textwidth]{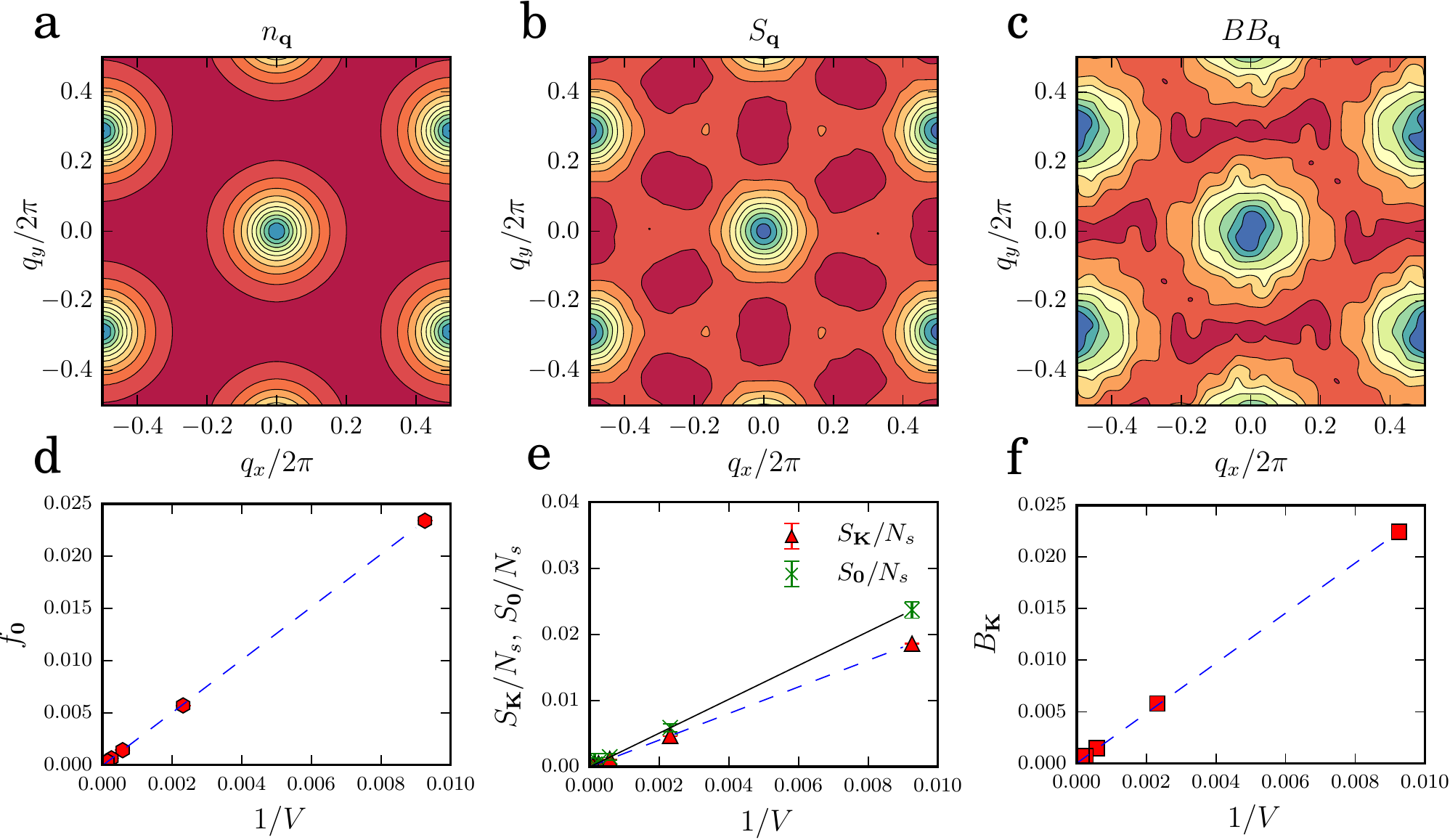}
\caption{ Off-diagonal $n_{\text{\bf q}}$ (a), diagonal  
  $S_{\text{\bf q}}$ (b), and bond $BB_{\text{\bf q}}$ (c)
structure factors for the $\text{W}_{\bigtriangleup}$ for a system 
with $N_{\text{s}}=24\times24$, $J_{z2}/J_{\pm}=4.5$, $J_{z1}/J_{\pm}=2$,  
and $T=J_{\pm}/24$. In each panel, the color scale represents the 
intensity of the structure factor. The corresponding finite-size 
scaling of select $\text{\bf q}$ values
in the structure factors $n_{\text{\bf 0}}$ (d) (red hexagons),  
$S_{\text{\bf 0}}$ and $S_{\text{\bf K}}$ (e) (red triangles and green, respectively), and
$B_{\text{\bf K}}$ (f) (red squares). 
Linear fits to the data in panels (d), (e), and (f) are 
represented as blue dashed and black solid lines.
\label{fig:sfactors} The zero-momentum peak of $BB_{\text{\bf 0}}$ 
in panel (c) has been removed. }
\end{figure*} 

The spin structure factors of the $\text{W}_{\bigtriangledown}$ phase 
are presented in Fig.(\ref{fig:sfactors}). We find none of the structure factors 
(off-diagonal $n_{\text{\bf q}}$ Fig.(\ref{fig:sfactors})a, diagonal  
$S_{\text{\bf q}}$ Fig.(\ref{fig:sfactors})b, and bond $BB_{\text{\bf q}}$ 
Fig.(\ref{fig:sfactors})c) displays peaks that signal the translational 
symmetry breaking. This is confirmed in the corresponding size scaling  
of selected peaks in Fig.(\ref{fig:sfactors})d through Fig.(\ref{fig:sfactors})e. 
For instance, $S_{\text{\bf K}}$ and $BB_{\text{\bf K}}$, which are 
non-vanishing in the VBS phase, quickly go to zero in the 
$\text{W}_{\bigtriangledown}$ phase, within the precision 
of our calculations. 

\begin{figure} 
  \includegraphics[width=0.44\textwidth]{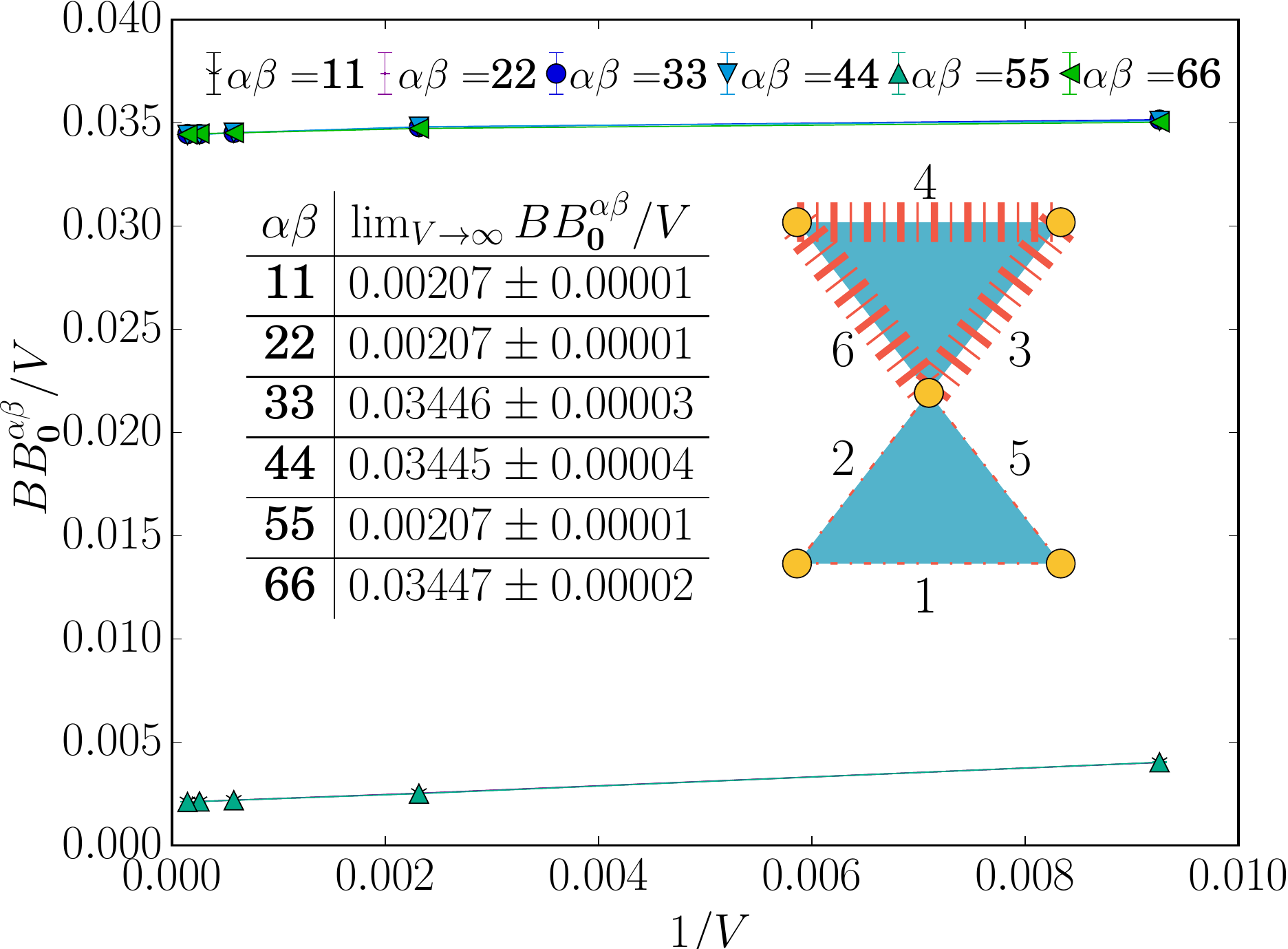}
  \caption{\label{fig:rotinv} The finite-size data for different sublattice 
  $BB^{\alpha\beta}_{\text{\bf 0}}$ as a function of $1/V$ in the featureless 
$\text{W}_{\bigtriangledown}$ phase at $J_{z2}/J_{\pm}=4.5$, $J_{z1}/J_{\pm}=2$. 
The insets show the extrapolated values 
$\lim_{V\to \infty }BB^{\alpha\beta}_{\text{\bf 0}}/V$ to the thermodynamic 
limit (table) and how this information translates into a real space pattern 
shown in the Kagome unit cell (drawing). }
\end{figure} 

We have removed the zero-momentum peak of the bond-bond 
structure factor since $B_{\text{\bf 0}}$ remains finite 
as $V\to\infty$ in all phases. However,
a closer look at the different sublattice $BB^{\alpha\beta}_{\text{\bf 0}}$ 
could reveal a potential rotational symmetry breaking in 
the unit cell of the Kagome lattice. In Fig.(\ref{fig:rotinv}) 
we present results for the finite-size scaling of 
$BB^{\alpha\beta}_{\text{\bf 0}}$ for the different bonds 
in the unit cell and their extrapolations to the thermodynamic limit. 
In the inset the extrapolations to the thermodynamic limit reveal that  
there is no symmetry breaking in the unit cell 
within the error bars of our simulation. 
In the hardcore boson language, the real-space pattern extracted 
from the extrapolations shows that 
the bosons are predominantly delocalized along the three sites of the 
strong triangles (1 boson per down triangle, in this case) 
since this pattern can be understood 
as the real space distribution of the square of the average 
kinetic energy along the bonds in the unit cell. 
A direct consequence of this observation is that
charge fluctuations in the weak triangle are larger 
than in the strong one. We quantify these fluctuations 
through a ``local'' uniform susceptibility 
\begin{equation}
\chi_{t}=3\beta\left[ \left \langle \left(\frac{1}{3}\sum_{i \in t}n_i\right)^2  \right \rangle  - \left \langle \left(\frac{1}{3}\sum_{i \in t}n_i\right)  \right\rangle^2\right], 
\end{equation}
where $n_i=S^{z}_i+1/2$. In particular, in the $\text{W}_{\bigtriangledown}$ 
phase for $J_{z2}/J_{\pm}=4.5$, $J_{z1}/J_{\pm}=2$, $\beta=J_{\pm}/60$, and 
$V=60\times60\times3$ we find that $\chi_{\bigtriangledown}=0.855\pm0.001$, 
whereas a significantly more fluctuating weak triangle $\chi_{\bigtriangleup}=7.386\pm0.002$ 
is found. Lastly, we examine the chiral-chiral 
correlation function $\langle E_{t} E_{t'} \rangle$, where 
$E_{t}={\mathbf S_{i_1}}\left( {\mathbf S_{ i_2}} 
\times {\mathbf S_{i_3}} \right)$ ($i_1,i_2,i_3 \in t$). Although the time reversal symmetry 
is explicitly broken by the finite magnetization in the model, chiral order may still potentially develop. 
In the limit where $J_{z2} \gg J_{z1}$, spin chirality may develop by correlating the fluctuations 
of the spins on the strong triangles as a way to offset the diagonal energy 
terms on the weak triangle. We find, however, that the chiral correlations 
are only enhanced as the $\text{W}_{\bigtriangledown}$ 
is approached, but the correlations still decay exponentially fast, 
as in the FM phase. In the appendix we demonstrate how to measure 
the chiral-chiral correlations in SSE. 

\section{\uppercase{Phase transitions}}

\subsection{FM-to-W phase transition}

\begin{figure} 
  \includegraphics[width=0.44\textwidth]{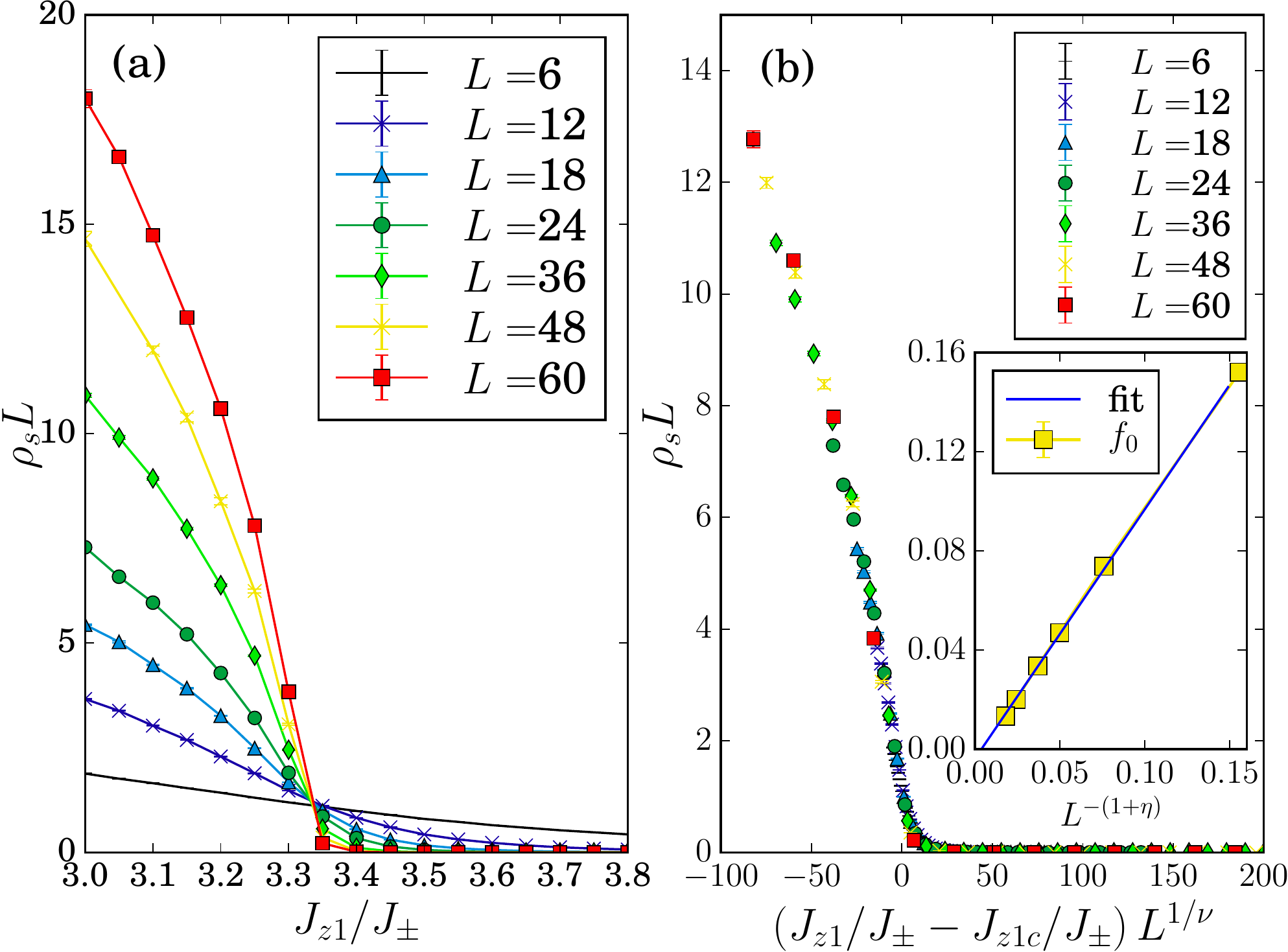}
\caption{\label{fig:XY} Finite-size scaling of the superfluid 
stiffness across the FM-to-W transition at fixed magnetization $m=-1/6$.
(a) The rescaled superfluid stiffness $\rho_sL$ as a function of $J_{z1}/J_{\pm}$. 
(b) The rescaled $\rho_sL$ as a function of $\left(J_{z1}/J_{\pm} - J_{z1c}/J_{\pm}  \right) L^{1/\nu}$.
The inset displays the condensate fraction $f_0$ as a function of $L^{-\left( \eta+1\right)}$. The blue line is a fit of the data to a straight line. 
To produce all three plots we have used the critical exponents $z=1$, $\nu=0.6717\pm 0.0001$, and $\eta=0.0381 \pm 0.0002$ obtained in Ref.\onlinecite{Campostrini2006}. 
We notice that the data in the inset are not produced with 
a canonical algorithm, but we emphasize that the average 
magnetization has been tuned to the canonical 
value within less than 0.1 percent error at the critical point.  }
\end{figure}

The phase diagram presented in Fig.\ref{fig:lattice}(a) 
exhibits three types of phase boundaries: 
FM-to-W, FM-to-VBS, and W-to-VBS phase transitions. We first 
examine the transition between the W and the FM phase. 
Since the tripartite entangled plaquette state W is a fully symmetric, 
then such a quantum critical line is expected to be a conventional $(d+1)$-dimensional 
order-disorder transition (with $d=2$). We consider two situations. First, 
the W phase is approached at strictly 
conserved magnetization. Second, the W phase is approached at a
fixed magnetic field $h_{\text{eff}}$. The transition at fixed 
magnetization is anticipated to be of the $(d+1)$-dimensional 
$O(2)$ vector model type.\cite{Fisher1989} 
If the W phase is approached at a fixed $h_{\text{eff}}$, the transition 
corresponds to the appearance of a dilute fluid of excess quasiparticles or 
holes on top of the W states where the magnetization deviates from $m=-1/6$, and is 
generically Gaussian.\cite{Fisher1989} For these types of continuous 
critical points in two dimensions, the superfluid stiffness 
scales as $\rho_s L^{z}=F_{\rho_s}\left( L^{1/\nu} 
\left( J-J_c \right),\beta / L^z \right)$,\cite{Fisher1973,Fisher1989,Sandvik2010} 
where $F_{\rho_S}$ is a scaling function, $z$ the dynamical 
critical exponent, $\nu$ the correlation length exponent, 
and $J-J_c$ the distance to the critical point in terms 
of the control parameter $J$. In Fig.\ref{fig:XY} we 
scrutinize the finite-size scaling of the superfluid stiffness 
at the transition between the FM and the $W_{\vartriangle}$ as 
a function of $J_{z1}/J_{\pm}$ at fixed $J_{z2}/J_{\pm}=2$, $m=-1/6$, 
and $\beta/L=1$. To produce the plots in 
Fig.\ref{fig:XY} we have used the critical exponents of 
$(d+1)$-dimensional $O(2)$ vector model, 
i.e., $z=1$, $\nu=0.6717\pm 0.0001$, and 
$\eta=0.0381 \pm 0.0002$.\cite{Campostrini2006} 
In Fig.\ref{fig:XY}(a) we show the rescaled superfluid 
stiffness $\rho_sL$ as a function of $J_{z1}/J_{\pm}$, 
which becomes system-size independent at the critical point 
$J_{z1c}/J_{\pm}=3.3325\pm0.0001$, as implied by the scaling
relation. We numerically extract $F_{\rho_s}$ 
by plotting the $\rho_sL$ as a function of $\left(J_{z1}/J_{\pm} 
- J_{z1c}/J_{\pm}  \right) L^{1/\nu}$, where a clear collapse is seen in 
Fig.\ref{fig:XY}(b). Finally, in the inset in Fig.\ref{fig:XY}(b), 
we analyze the size scaling of the condensate fraction $f_0$,
which vanishes as $f_0\sim L^{-\left( \eta+1\right)}$ 
at the critical point.\cite{Carrasquilla2012} We find that our data 
for $f_0$ are well described by a 
straight line when plot as a function of $L^{-\left( \eta+1\right)}$ 
with the $\eta$ obtained in Ref.\onlinecite{Campostrini2006}. 
Our results are consistent with a transition described 
by the $(d+1)$-dimensional $O(2)$ vector model. 

\begin{figure} 
  \includegraphics[width=0.44\textwidth]{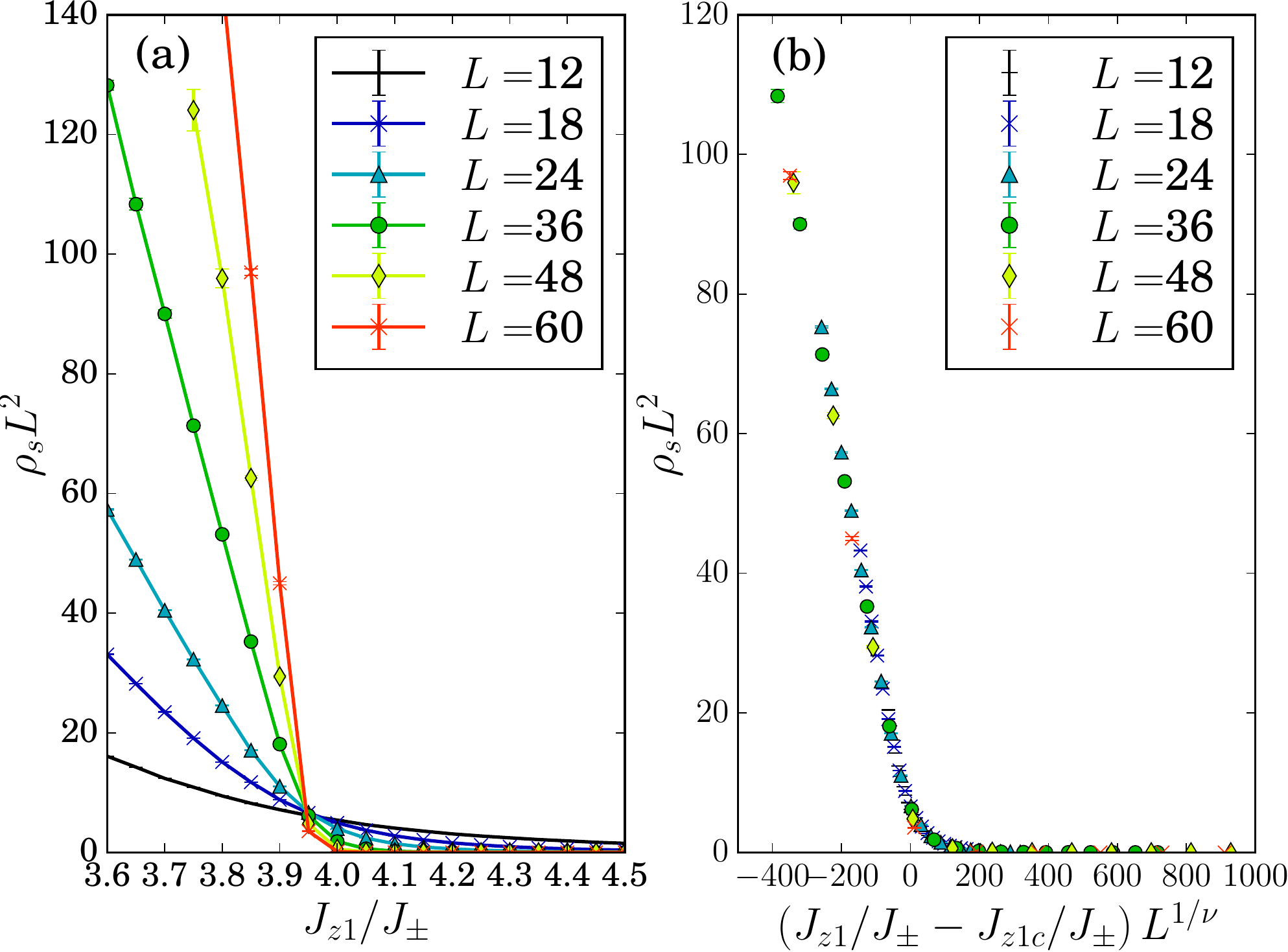}
  \caption{\label{fig:gaussian} Finite-size scaling of the superfluid stiffness 
  across the FM-to-W transition at fixed magnetic field $h_{\text{eff}}/J_{\pm}=-2.935$, 
$J_{z2}/J_{\pm}=2$, and $\beta/L^z=0.1$.
(a) The rescaled superfluid stiffness $\rho_sL^2$ as a function of $J_{z1}/J_{\pm}$. 
(b) The rescaled $\rho_sL^2$ as a function of 
$\left(J_{z1}/J_{\pm} - J_{z1c}/J_{\pm}  \right) L^{1/\nu}$. 
To produce these two plots we have used the critical exponents 
$z=2$, $\nu=0.5$, and $J_{z1c}/J_{\pm}=3.9428\pm0.0004$. }
\end{figure}

Similarly, we investigate the generic transition at 
fixed $h_{\text{eff}}/J_{\pm}=-2.935$, $J_{z2}/J_{\pm}=2$, 
and $\beta/L^z=0.1$. The critical exponents that characterize 
this transition are $z=2$, $\eta=0.0$, $\nu=0.5,$\cite{Fisher1989} 
which we use in the following to locate the generic 
critical point. In Fig.\ref{fig:gaussian}(a) we show the 
rescaled superfluid stiffness $\rho_sL^2$ as a function 
of $J_{z1}/J_{\pm}$, which becomes system-size independent 
at the critical point $J_{z1c}/J_{\pm}=3.9428\pm0.0004$.  
In Fig.\ref{fig:gaussian}(b) we extract the scaling function 
$F_{\rho_s}$ across the Gaussian critical point, where we 
observe again a clear collapse to a unique curve. 
 
In summary, the phase transitions from the FM toward the W 
states are consistent with the picture of $(d+1)$-dimensional 
order-disorder transitions and with the picture 
we have described in Section~\ref{Winsul}, namely, that the W 
states are insulators adiabatically connected to 
product states of tripartite entangled plaquette states.

\subsection{FM-to-VBS transition}
\begin{figure*}[t]
\includegraphics[trim=0 0 0 0, clip,width=0.8\textwidth]{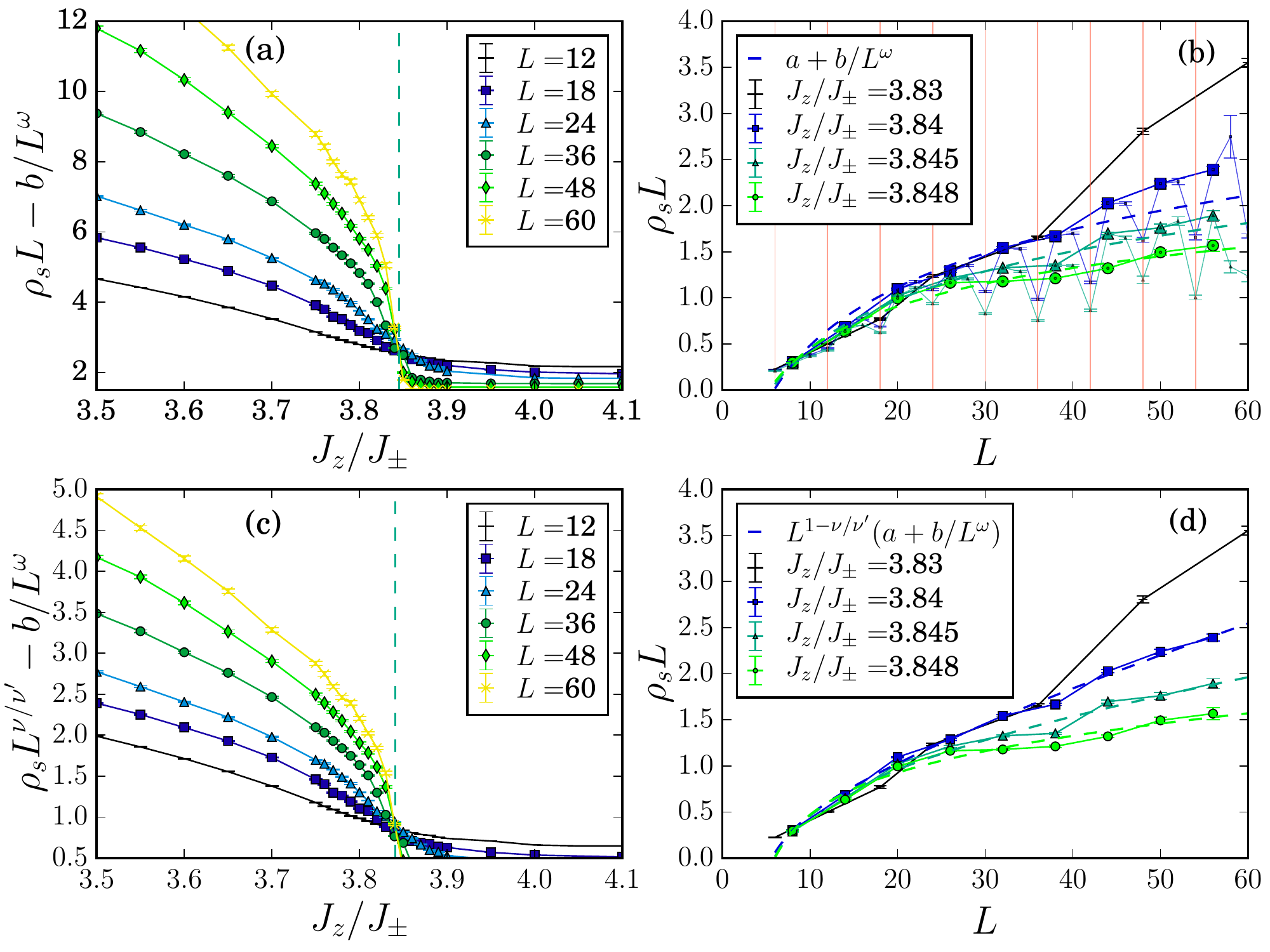}
  \caption{\label{fig:FMtoVBS} Finite-size scaling of the 
  superfluid stiffness across the FM-to-VBS transition  
at fixed magnetization $m_z=-1/6$ and $\beta/L=1$. A conventional continuous  scaling is presented in (a) and (b). (a) 
The rescaled superfluid stiffness $\rho_sL-bL^{-\omega}$ as a function of $J_{z}/J_{\pm}$ for different system sizes. (b) 
The scaled superfluid stiffness $\rho_s L$ as a function of the system size $L$ for several values of $J_{z}/J_{\pm}$ 
near the critical point (solid lines). Fits to the scaling form $\rho_s L=a+bL^{-\omega}$ (dashed lines). The fits are obtained
using only observations following the values of $L$ which accommodate the wave vectors $\text{\bf q}=(2\pi/3,0)$ (displayed with larger
symbols and signaled by thin vertical lines ). A two-length scale scenario is explored in (c) and (d). (c) The rescaled superfluid stiffness 
$\rho_sL^{\nu/\nu^{\prime}}-bL^{-\omega}$ as a function of $J_{z}/J_{\pm}$ for different system sizes. (d) The scaled superfluid stiffness 
$\rho_s L$ as a function of the system size $L$ for several values of $J_{z}/J_{\pm}$ near the critical point (solid lines). Fits to the scaling form 
$\rho_s L=L^{1-\nu/\nu^{\prime}}\left(a+bL^{-\omega}\right)$ (dashed lines). The fits are obtained using only observabtions following 
the values of $L$ which accommodate the wave vectors $\text{\bf q}=(2\pi/3,0)$.  } 

\end{figure*}

We now turn our attention to the FM-to-VBS quantum phase transition. This transition has 
been numerically studied in Ref.\onlinecite{Isakov2006} and Ref.\onlinecite{Damle2006} 
using grand-canonical SSE Monte Carlo. The authors in Ref.\onlinecite{Damle2006} reported 
evidence of a continuous phase transition based on data for the Binder cumulant and the 
superfluid stiffness consistent with a non-Ginzburg-Landau deconfined critical point, 
but did not rule out the possibility of a weak first-order phase transition. On the other 
hand, Ref.\onlinecite{Isakov2006} reported  finite-size scaling of the superfluid stiffness, 
structure factor, and kinetic energy histograms on larger system sizes. The authors found 
evidence of a first-order quantum phase transition. Their strongest evidence in favor of a 
first-order transition was based on extremely low-temperature histograms of the kinetic energy 
which exhibited a double peak structure signaling coexistence at the critical point. Here we 
revisit this critical point using canonical measurements of the superfluid stiffness and 
structure factor supplemented with several finite-size scaling analyses. First, we examine 
the superfluid stiffness. Assuming a continuous phase transition where $z=1$, we compute the 
rescaled superfluid stiffness $\rho_sL$ as a function of $J_{z2}/J_{\pm}=J_{z1}/J_{\pm}=J_z/J_{\pm}$ with $\beta/L=1$ and 
explore two finite-size scaling scenarios. First, we consider a conventional scaling scenario
described by a divergent length scale $\xi\propto \delta^{-\nu}$, where $\delta=J_z-J_{zc}$ controls
the distance to the quantum critical point and $\nu$ is the correlation length exponent. Assuming 
a dynamical exponent $z=1$, for a system of linear size $L$ close to $\delta=0$ the superfluid stiffness 
is singular and scales as $\rho_s (\delta,L)=L^{-z}F_{\rho_s}\left( \delta L^{1/\nu},L^{-\omega} \right)$, 
where $\omega$ is a correction-to-scaling exponent. This relation means that at the critical point  
$\rho_s L=a+bL^{-\omega}$ (with $a$ and $b$ constants), and that  $\rho_s L-bL^{-\omega}$ becomes 
system-size independent. In Fig.\ref{fig:FMtoVBS}(a) we plot the relation $\rho_s L-bL^{-\omega}$ vs
$J_z/J_{\pm}$ for our numerical estimates of $\rho_s$, which become approximately system-size 
independent around $J_z/J_{\pm}=3.845\pm0.004$. Note that the crossing point tends to move slowly 
towards lower values of $J_z/J_{\pm}$ and that the value of $\rho_sL\equiv\rho_s^cL$ between two subsequent system 
sizes is slowly diverging with the system size, which means that corrections to scaling are significant. 
A more detailed picture arises by considering the size scaling around the critical point: in Fig.\ref{fig:FMtoVBS}(b) we 
plot $\rho_s L$ vs linear system size $L$ for different values of $J_z/J_{\pm}$ near the critical point. On 
increasing $J_z/J_{\pm}$, we notice the development of strong system-size dependent oscillations in 
the superfluid stiffness which become stronger as the VBS phase is approached. The period of the oscillations 
can be traced back to the translational symmetry breaking of the VBS phase since the minima of the $\rho_s$ 
oscillations appear at system sizes that exactly accommodate the wave vector of the VBS pattern, i.e., 
$\text{\bf q}=\text{\bf K}=(2\pi/3,0)$ (see thin vertical lines in Fig.\ref{fig:FMtoVBS}(b)). Because of the 
oscillations, fitting the data (to either all the data or to just the local minima or  maxima ) to the scaling 
form $\rho_s L=a+bL^{-\omega}$ produce estimates for $a,b$ and $\omega$ with error bars in the first significant digits. 
Apart from the conventional possibilities discussed above, the slow divergence of the stiffness near the critical point opens 
up the possibility for yet another scenario, i.e., one with a continuous transition with two diverging length scales, 
as recently proposed in Ref.~\onlinecite{Shao213}. A prediction from the two-length scale scenario is that if there is 
a second large-$L$  scale controlled by $\delta L^{1/\nu^{\prime}}$, the superfluid stiffness behaves as 
$\rho_s L=L^{1-\nu/\nu^{\prime}}\left (a+bL^{-\omega}\right)$ at the critical point,\cite{Shao213} which we now test.  
In Fig.\ref{fig:FMtoVBS}(c) we display $\rho_s L^{z\nu/\nu'}-b/L^{\omega}$, which, once again,  becomes approximately 
system-size independent near the estimate of the critical point $J_z/J_{\pm}=3.845\pm0.004$. The parameters used in 
Fig.\ref{fig:FMtoVBS}(c) are obtained from fitting our data for $\rho_s\left( L, J_z/J_{\pm} \right)$ to the form predicted by 
the two-length scale scenario $\rho_s L=L^{1-\nu/\nu^{\prime}}\left (a+bL^{-\omega}\right)$ presented in Fig.\ref{fig:FMtoVBS}(d).
We find that $\nu/\nu^{\prime}\approx0.4\pm0.2$, while $\omega\approx 2 \pm 2$. Even though the quality of the fits to the 
two-length scaling forms is notably better than the conventional continuous scaling, the significance of some of the fitting 
parameters we obtain is again compromised by the large error bars resulting from the oscillations in the data. 

\begin{figure} 
  \includegraphics[width=0.44\textwidth]{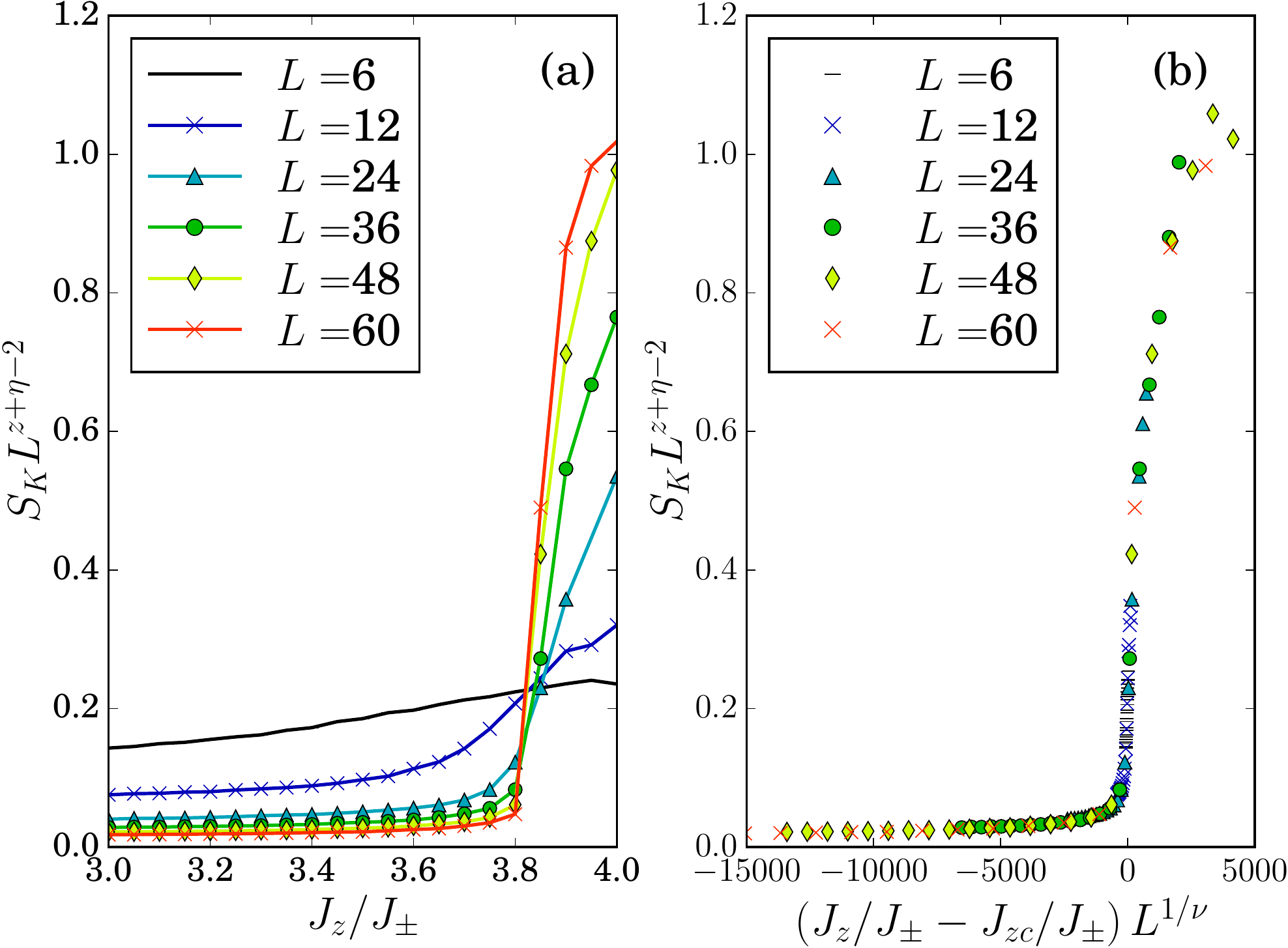}
\caption{\label{fig:SKFMtoVBS} Finite-size scaling of the structure 
factor $S_{\text{\bf K}}$ across the FM-to-VBS transition as a function of $J_z/J_{\pm}$
at fixed magnetization $m_z=-1/6$ and $\beta/L=1$.
(a) The rescaled structure factor $S_{\text{\bf K}}L^{z+\eta-2}$  as a function of $J_{z}/J_{\pm}$. 
(b) The structure factor $S_{\text{\bf K}}$ as a function of the resulting  $(J_{z}/J_{\pm} -J_{zc}/J_{\pm})L^{1/\nu}$ across the phase transition. }
\end{figure}

We also discuss the finite-size scaling of the structure 
factor $S_{\text{\bf q}}$ at momentum $\text{\bf q}=
\text{\bf K}=(2\pi/3,0)$, which scales to a finite value in
the thermodynamic limit inside the VBS phase and to 
zero in the FM phase. Assuming a continuous phase 
transition, the structure factor scales as
$S_{\text{\bf K}}L^{z+\eta-2}=F_{S}\left(   \left( J-J_c    
\right)L^{1/\nu},\beta/L^z \right)$.\cite{Isakov2006} 
In Fig.\ref{fig:SKFMtoVBS} we analyze numerical data
for $S_{\text{\bf K}}$ using size scaling. Assuming $z=1$,  
Fig.\ref{fig:SKFMtoVBS}(a) displays the rescaled 
$S_{\text{\bf K}}L^{z+\eta-2}$ vs $J_{z}/J_{\pm}$, while in
Fig.\ref{fig:SKFMtoVBS}(b) we attempt at numerically 
obtaining the scaling function $F_{S}$ by plotting 
$S_{\text{\bf K}}L^{z+\eta-2}$ vs $(J_{z}/J_{\pm} 
-J_{zc}/J_{\pm})L^{1/\nu}$. To produce the collapse 
in Fig.\ref{fig:SKFMtoVBS}(b) we find that $\eta=0.10\pm0.01$ and 
$\nu=0.40\pm0.02$. Even though the collapse looks 
compellingly consistent with criticality, the absence of 
a crossing in Fig.\ref{fig:SKFMtoVBS}(a) and the 
appearance of a jump in $S_{\text{\bf K}}L^{z+\eta-2}$ 
suggests that the transition may indeed be first order. We finalize our analysis by 
mentioning that we have also performed simulations 
(not shown) for the same transition across the 
$J_{z2}/J_{\pm}=J_{z1}/J_{\pm}-0.18$ line reaching the 
same conclusions. 

To conclude this section, we have reexamined the FM-to-VBS 
transition using a canonical algorithm with which we 
obtain data for the superfluid stiffness and structure factor 
near the critical point. Apart from the conventional scenarios, and since this transition may potentially 
exhibit non-classical behavior with the presence of fractionalized 
excitations, we have also explored the possibility of a scaling 
with two length scales, which has been recently introduced in 
Ref.~\onlinecite{Shao213} in the context of a $J-Q$ model exhibiting 
such phenomenon. Although our data seem consistent with that 
scenario, the large error bars associated with fitting numerical data to such 
predictions highlight the complications inherent to the application of this type 
of analysis to models like ours (Eq.~\ref{HAM}). Furthermore, the 
scaling of the structure factor, which shows drifting crossings between system 
sizes, is perhaps consistent with the idea that this transition is first order as suggested 
in Ref.\onlinecite{Isakov2006}. Our numerical estimate for the critical point 
based on the canonical measurements $J_{zc}/J_{\pm}=3.845\pm0.004$ is slightly 
below the previous grand-canonical results\cite{Isakov2006} $J_{zc}/J_{\pm}=3.898\pm0.001$.

\section{VBS-to-W transition}

\begin{figure} 
  \includegraphics[width=0.44\textwidth]{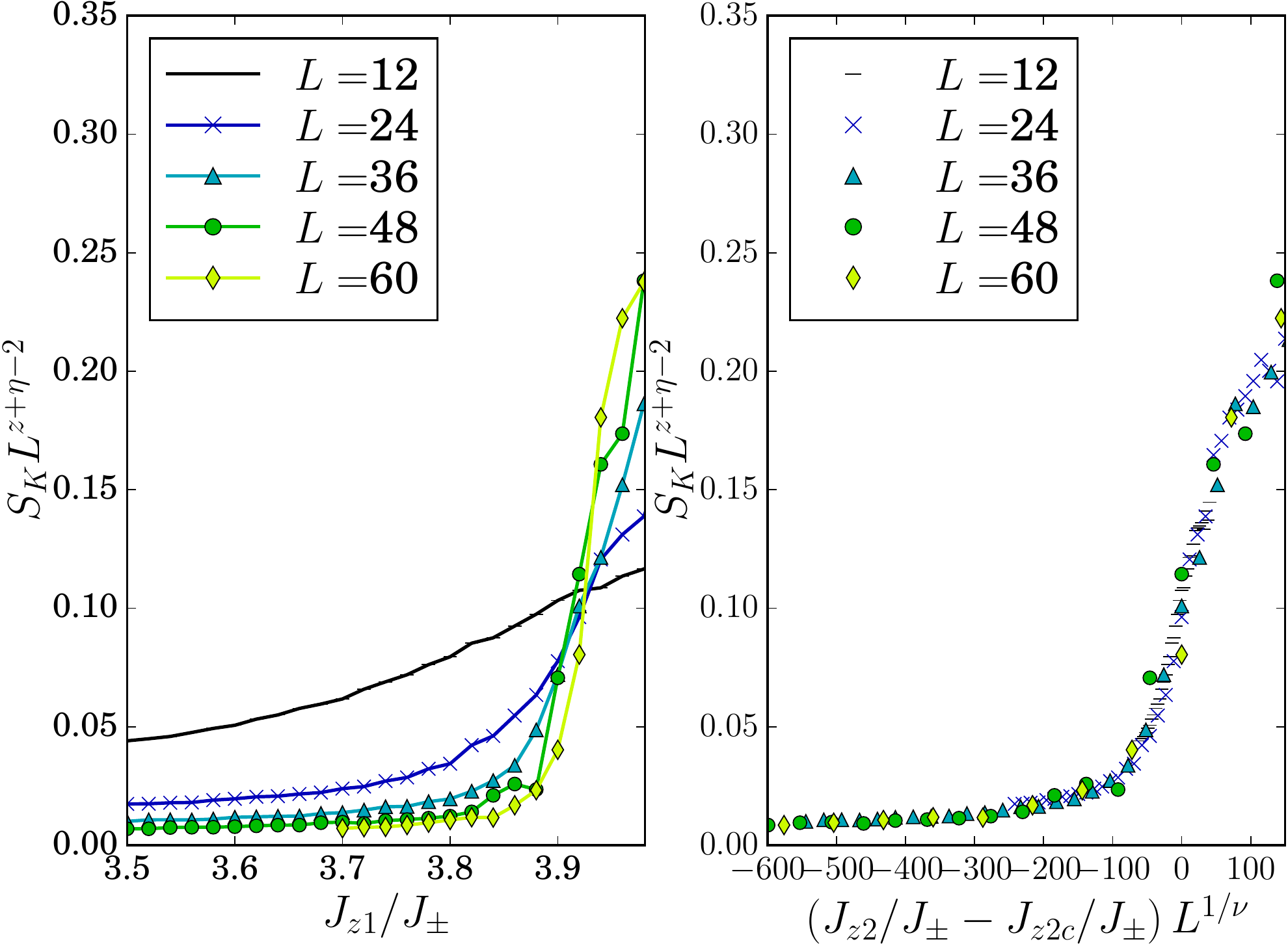}
  \caption{\label{fig:WtoVBS} 
Finite-size scaling of the structure factor $S_{\text{\bf K}}$ across the VBS-to-W transition as a function of $J_{z2}/J_{\pm}$
at fixed magnetization $m_z=-1/6$ and $\beta/L=1$.
(a) The rescaled structure factor $S_{\text{\bf K}}L^{z+\eta-2}$  as a function of $J_{z}/J_{\pm}$.
(b) The structure factor $S_{\text{\bf K}}$ as a function of the resulting  $(J_{z}/J_{\pm} -J_{zc}/J_{\pm})L^{1/\nu}$ across the phase transition.}
\end{figure}

\begin{figure} 
  \includegraphics[width=0.44\textwidth]{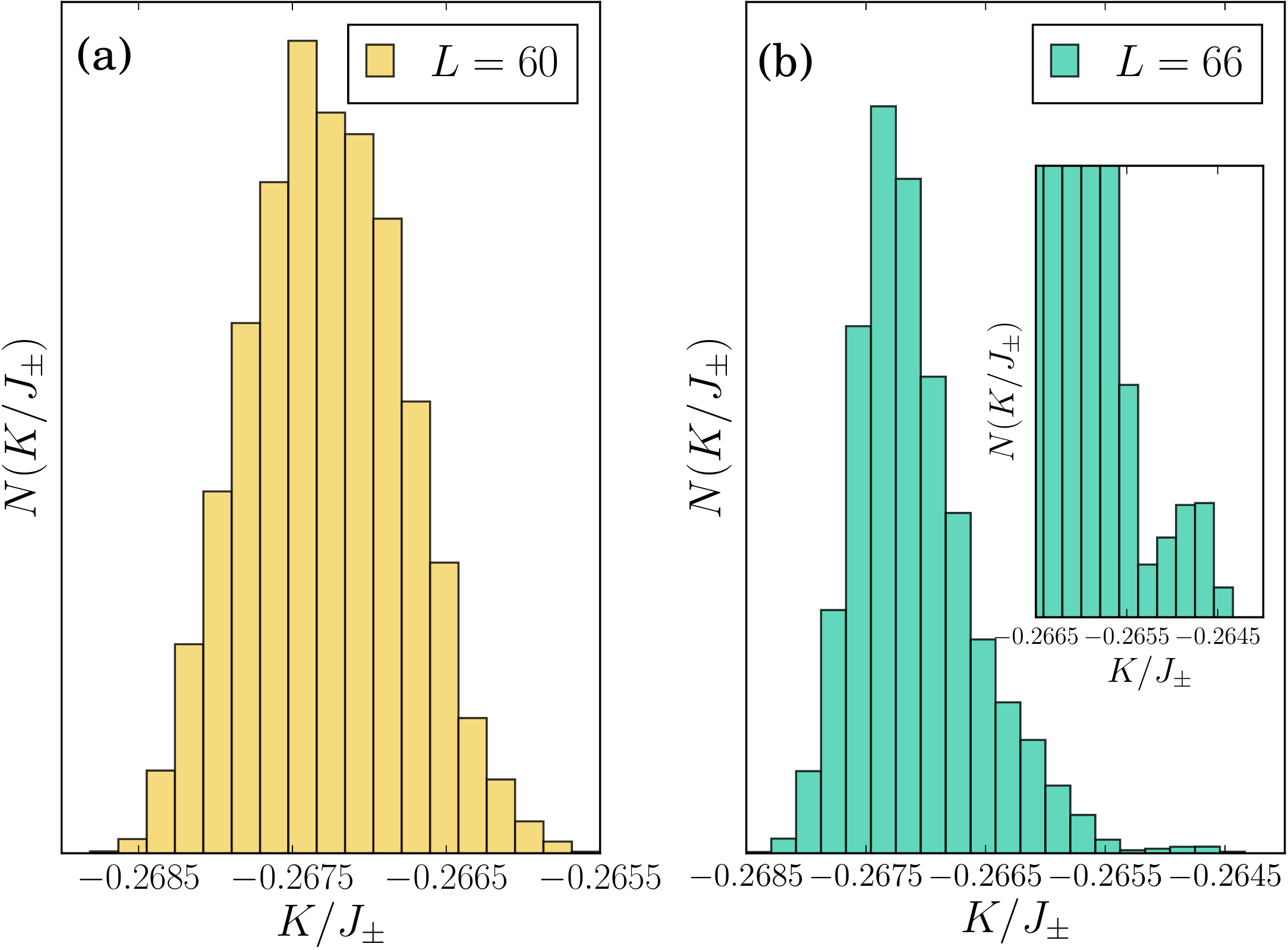}
  \caption{\label{fig:histo} Kinetic energy histograms for system sizes $L=60$ 
  (a) and  $L=66$ (b) near the critical point. The inset 
zooms in the smaller peak for $L=66$. }
\end{figure}

The phase diagram presented in Fig.\ref{fig:lattice}(a) 
features a VBS-to-W-state quantum phase transition which we now inspect.  The VBS present in our model
is characterized by a three-fold degenerate ground state where translational symmetry is broken and signaled by the formation of Bragg peaks in the diagonal and 
bond structure factors at $\text{\bf q}=\text{\bf K}=(2\pi/3,0)$. We thus expect that the related $(d+1)$ classical model showing the same universality class 
is the three-dimensional three-state, $Z_3$, clock model, which is equivalent to the three-dimensional three-state Potts model.\cite{Wu1982,Bonnes2011} This model 
is known to exhibit a first-order phase transition.\cite{Wu1982,Herrmann1979} Thus we anticipate that the VBS-to-W transition is first order. 
In Fig.\ref{fig:WtoVBS} we show finite-size scaling of the diagonal structure factor $S_{\text{\bf K}}$ across the VBS-to-W transition. Assuming 
$z=1$, in Fig.\ref{fig:WtoVBS}(a) we examine the rescaled $S_{\text{\bf K}}L^{z+\eta-2}$ as a function of  $J_{z2}/J_{\pm}$ at fixed $J_{z1}/J_{\pm}=4.5$, 
$\beta/L=1$, and $m_z=-1/6$. In Fig.\ref{fig:WtoVBS}(b) we attempt at obtaining the the scaling function $F_{S}$ by plotting $S_{\text{\bf K}}L^{z+\eta-2}$ vs 
$(J_{z2}/J_{\pm} -J_{z2c}/J_{\pm})L^{1/\nu}$. In order to produce Fig.\ref{fig:SKFMtoVBS} we have used $\eta=-0.30\pm0.01$ and $\nu=0.50\pm 0.04$. 
Even though the data collapse may appear consistent with criticality, we 
argue that the absence of a clear crossing in Fig.\ref{fig:SKFMtoVBS}(a) and the 
unusual negative value of $\eta$ suggests that the transition is instead first order, in agreement with the expected three-dimensional three-state Potts model. 
Furthermore, we consider histograms of the order parameter $S_{\text{\bf K}}$, the total energy, and the kinetic energy at the critical point. We find that 
the histograms of the energy and order parameter are not bimodal for all system sizes accessible in our simulations (not shown). Instead, in Fig.\ref{fig:histo} 
we show histograms of the kinetic energy $K/J_{\pm}$ for system sizes $L=60$ (Fig.\ref{fig:histo}(a)) and $L=66$ (Fig.\ref{fig:histo}(b)) where we find that 
for the largest system we could simulate ($L=66$) the kinetic energy starts developing a two-peak histogram with a dominant beside a small, albeit statistically 
robust, peak suggesting the onset of phase coexistence. While this signal seems rather stable, definitive evidence in favor of a first-order phase transition 
requires verifying that the double-peak histograms remain stable for larger system sizes unavailable in our simulation setup. 

\section{Conclusion}

Inspired by various molybdenum-based cluster magnets,~\cite{Sheckelton2012,Mourigal2014,Sheckelton2014,Haraguchi2015,Chen2014,Chen2015}
we have studied the phase diagram of a generic XXZ spin model 
on the anisotropic Kagome lattice
using large-scale SSE quantum Monte Carlo simulations. 
We have found a remarkable 
tripartite entangled plaquette state surrounded by a valence-bond solid state 
and a ferromagnet state, and studied the respective transitions between these phases. We find that 
all the transitions toward the tripartite entangled plaquette state are conventional order-disorder
transitions, either continuous (FM-W) or first order (VBS-W), which supports the idea that 
the W phase is a featureless symmetric state. 

We have reexamined the FM-to-VBS transition using a canonical algorithm in the light of a recently proposed 
scaling analysis with two length scales.\cite{Shao213} While our data seem consistent with that scenario, the 
large error bars associated with fitting numerical data to the predictions of the two-length scaling analysis prevented 
us from drawing a firm conclusion about the applicability of such scenario to our data. Furthermore, the
scaling of the structure factor, which shows drifting crossings between system
sizes, is perhaps consistent with the idea that this transition is first order as suggested
in Ref.\onlinecite{Isakov2006}. 

As we described in Sec.~\ref{sec2}, the XXZ spin model in our work describes 
the charge sector physics of the cluster magnet, and, as such,  
one expects a one-to-one mapping between the phases in our generic phase 
diagram (Fig.\ref{fig:lattice}) and the phases of the extended 
Hubbard model used in their description. In particular, the FM phase 
in our phase diagram corresponds to the Fermi liquid phase, 
the tripartite entangled plaquette state W corresponds to the cluster 
Mott insulator with one electron localized on every {\sl strong} triangle, 
and the VBS phase corresponds to the plaquette charge ordered state. 

Besides the relevance to cluster magnets, our results directly 
apply to other areas outside of condensed matter. In
particular, strongly correlated bosonic atoms can be 
loaded onto highly tunable {\it trimerized} 
optical Kagome lattices.\cite{Santos2004,Damski2005,Gyu-Boong2012} 
Such systems are realizations of the XXZ model discussed here
via the mapping between spin-1/2 
and hardcore boson models.\cite{Matsuda01011957}

\acknowledgements
We would like to thank G. Baskaran, Y. Qi, S. Sachdev, 
M. Stoudenmire, Y. Wan, and W. Witczak-Krempa for enlightening discussions. 
We thank F. Hanington for a careful reading of the manuscript. 
We are especially indebted to A. Sandvik for discussions and suggestions 
related to the scaling forms with two divergent length scales.
This research was supported by NSERC of Canada, the Perimeter 
Institute for Theoretical Physics, and the John Templeton
Foundation. G.C. acknowledges support from the Thousand-Youth-Talent 
program of People's Republic of China. R.G.M.~acknowledges support 
from a Canada Research Chair. Research at Perimeter Institute is supported 
through Industry Canada and by the Province of Ontario through the Ministry 
of Research \& Innovation. Numerical simulations were carried out on the 
Shared Hierarchical Academic Research Computing Network (SHARCNET).

\section*{appendix}

\begin{figure} [ht]
  \includegraphics[width=0.44\textwidth]{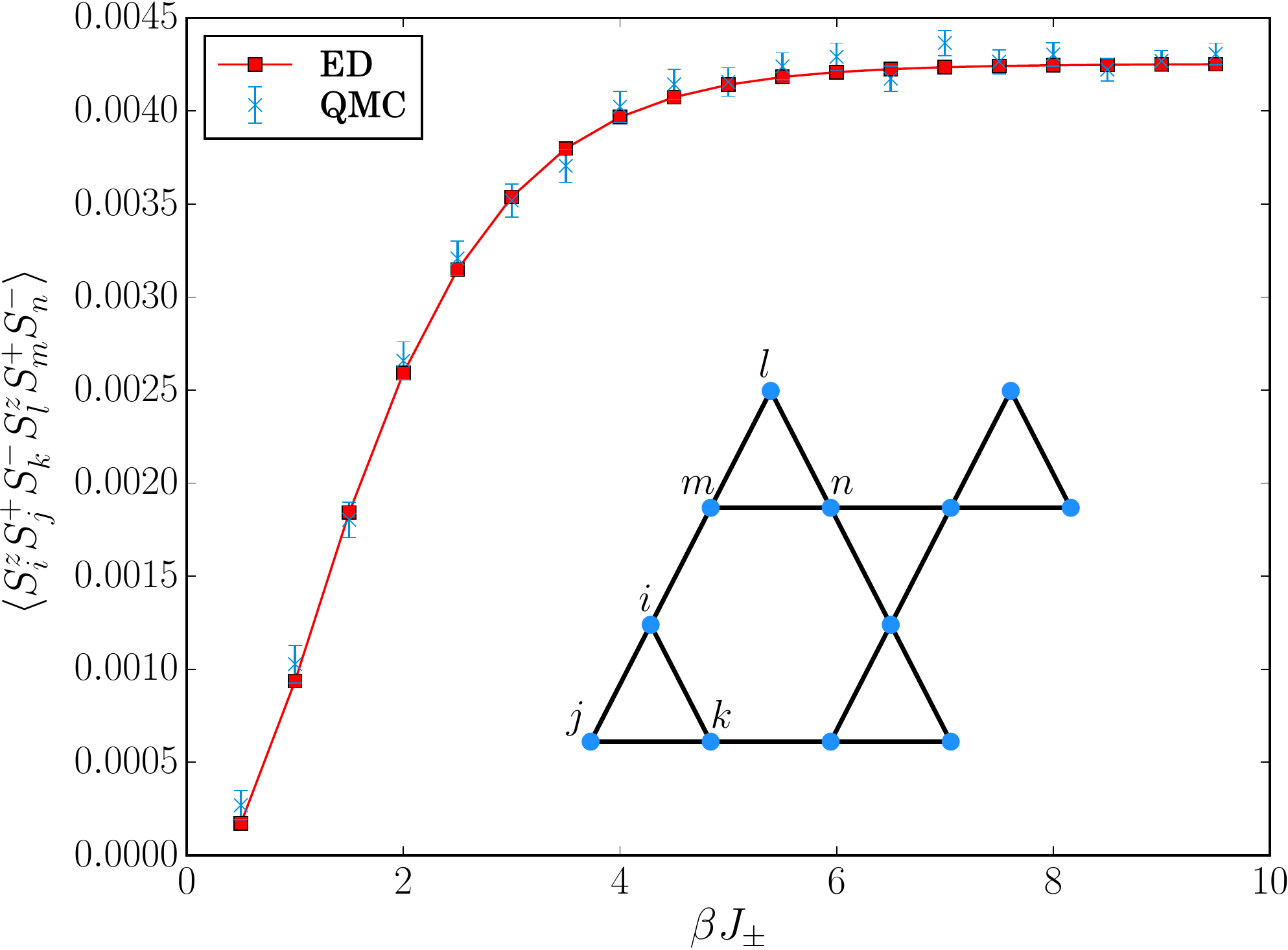}
  \caption{\label{fig:chiral} The $\langle S^{z}_{i} S^{+}_{j} S^{-}_{k} S^{z}_{l} S^{+}_{m} S^{-}_{n} \rangle$ correlation as a function of the inverse 
temperature $\beta J_{\pm}$ on a $V=2\times2\times3$ lattice with periodic boundary conditions at $h_{\text{eff}}/J_{\pm}=-2.935$,
$J_{z2}/J_{\pm}=2$, and $J_{z1}/J_{\pm}=4.2$ using quantum Monte Carlo (QMC) and exact diagonalization (ED). }

\end{figure}

In this section we briefly detail the procedure to measure chiral-chiral correlation functions in SSE quantum Monte Carlo simulations. Measurements of chiral-chiral correlation 
functions defined as $\langle E_{t} E_{t'} \rangle$, where $E_{t}={\mathbf S_{i_1}}\left( {\mathbf S_{ i_2}}\times {\mathbf S_{i_3}} \right)$ ($i_1,i_2,i_3 \in t$),
are possible within SSE whenever the basic Hamiltonian breakup contains the triangles defined by the different indices ($i_1,i_2,i_3 \in t$) in the correlators. Since 
the Hamiltonian breakup we have used in our study uses the underlying corner-sharing triangles\cite{Melko2007} defining the Kagome lattice,  chiral-chiral 
correlations defined over those triangles can be naturally obtained within our simulations efficiently. By expanding the vector products in the 
correlator $\langle E_{t} E_{t'} \rangle$, the SSE measurements are simplified to a combination of terms the form 
$\langle S^{z}_{i} S^{+}_{j} S^{-}_{k} S^{z}_{i'} S^{+}_{j'} S^{-}_{k'} \rangle$ where $i,j,k\in t$ and $i',j',k'\in t'$. Terms of the form above can be measured through 
the estimator 
\begin{eqnarray}
&&\langle S^{z}_{i} S^{+}_{j} S^{-}_{k} S^{z}_{i'} S^{+}_{j'} S^{-}_{k'} \rangle 
\nonumber 
\\
&&\quad\quad =\frac{4}{\left( \beta J_{\pm} \right)^2} \left \langle \left(n-1\right) \sum_t S^{z}_i[t] S^{z}_{i'}[t_{\text{next}}] \right \rangle,
\label{chi}
\end{eqnarray}
where the sum runs for ordered sub-sequences and the operators $S^{+}_{j} S^{-}_{k}$ appear at imaginary-time slice $t$ followed by a 
$S^{+}_{j'} S^{-}_{k'}$ at $t_{\text{next}}$ in  the operator sequence $S_n$, and $n$ is the expansion order.\cite{sandvik1999,Melko2007} We have implemented the 
estimator in Eq.\ref{chi} for each of the terms appearing in the chiral-chiral correlation function. To test the validity of our approach, in Fig.\ref{fig:chiral} 
we benchmark SSE estimates of one term, $\langle S^{z}_{i} S^{+}_{j} S^{-}_{k} S^{z}_{l} S^{+}_{m} S^{-}_{n} \rangle$, on a small $V=2\times2\times3$ cluster with 
periodic boundary conditions against exact diagonalization (ED) calculations. The sites $i,j,k$ and $l,m,n$ are depicted in the inset. The agreement of our SSE 
calculations with the ED results validates our approach.

\vspace{2cm}


\begin{thebibliography}{45}
\expandafter\ifx\csname natexlab\endcsname\relax\def\natexlab#1{#1}\fi
\expandafter\ifx\csname bibnamefont\endcsname\relax
  \def\bibnamefont#1{#1}\fi
\expandafter\ifx\csname bibfnamefont\endcsname\relax
  \def\bibfnamefont#1{#1}\fi
\expandafter\ifx\csname citenamefont\endcsname\relax
  \def\citenamefont#1{#1}\fi
\expandafter\ifx\csname url\endcsname\relax
  \def\url#1{\texttt{#1}}\fi
\expandafter\ifx\csname urlprefix\endcsname\relax\def\urlprefix{URL }\fi
\providecommand{\bibinfo}[2]{#2}
\providecommand{\eprint}[2][]{\url{#2}}

\bibitem[{\citenamefont{Tamura et~al.}(2006)\citenamefont{Tamura, Nakao, and
  Kato}}]{Tamura2006}
\bibinfo{author}{\bibfnamefont{M.}~\bibnamefont{Tamura}},
  \bibinfo{author}{\bibfnamefont{A.}~\bibnamefont{Nakao}}, \bibnamefont{and}
  \bibinfo{author}{\bibfnamefont{R.}~\bibnamefont{Kato}},
  \bibinfo{journal}{Journal of the Physical Society of Japan}
  \textbf{\bibinfo{volume}{75}}, \bibinfo{pages}{093701}
  (\bibinfo{year}{2006}).

\bibitem[{\citenamefont{Matan et~al.}(2010)\citenamefont{Matan, Ono, Fukumoto,
  Sato, Yamaura, Yano, Morita, and Tanaka}}]{Matan2010}
\bibinfo{author}{\bibfnamefont{K.}~\bibnamefont{Matan}},
  \bibinfo{author}{\bibfnamefont{T.}~\bibnamefont{Ono}},
  \bibinfo{author}{\bibfnamefont{Y.}~\bibnamefont{Fukumoto}},
  \bibinfo{author}{\bibfnamefont{T.~J.} \bibnamefont{Sato}},
  \bibinfo{author}{\bibfnamefont{J.}~\bibnamefont{Yamaura}},
  \bibinfo{author}{\bibfnamefont{M.}~\bibnamefont{Yano}},
  \bibinfo{author}{\bibfnamefont{K.}~\bibnamefont{Morita}}, \bibnamefont{and}
  \bibinfo{author}{\bibfnamefont{H.}~\bibnamefont{Tanaka}},
  \bibinfo{journal}{Nat Phys} \textbf{\bibinfo{volume}{6}},
  \bibinfo{pages}{865} (\bibinfo{year}{2010}).

\bibitem[{\citenamefont{Sachdev}(1999)}]{Sachdev2011}
\bibinfo{author}{\bibfnamefont{S.}~\bibnamefont{Sachdev}},
  \emph{\bibinfo{title}{Quantum Phase Transitions}}
  (\bibinfo{publisher}{Cambridge University Press},
  \bibinfo{address}{Cambridge, UK}, \bibinfo{year}{1999}).

\bibitem[{\citenamefont{Vasiliev et~al.}(2015)\citenamefont{Vasiliev, Volkova,
  Zvereva, Koshelev, Urusov, Chareev, Petkov, Sukhanov, Rahaman, and
  Saha-Dasgupta}}]{Vasiliev2015}
\bibinfo{author}{\bibfnamefont{A.~N.} \bibnamefont{Vasiliev}},
  \bibinfo{author}{\bibfnamefont{O.~S.} \bibnamefont{Volkova}},
  \bibinfo{author}{\bibfnamefont{E.~A.} \bibnamefont{Zvereva}},
  \bibinfo{author}{\bibfnamefont{A.~V.} \bibnamefont{Koshelev}},
  \bibinfo{author}{\bibfnamefont{V.~S.} \bibnamefont{Urusov}},
  \bibinfo{author}{\bibfnamefont{D.~A.} \bibnamefont{Chareev}},
  \bibinfo{author}{\bibfnamefont{V.~I.} \bibnamefont{Petkov}},
  \bibinfo{author}{\bibfnamefont{M.~V.} \bibnamefont{Sukhanov}},
  \bibinfo{author}{\bibfnamefont{B.}~\bibnamefont{Rahaman}}, \bibnamefont{and}
  \bibinfo{author}{\bibfnamefont{T.}~\bibnamefont{Saha-Dasgupta}},
  \bibinfo{journal}{Phys. Rev. B} \textbf{\bibinfo{volume}{91}},
  \bibinfo{pages}{144406} (\bibinfo{year}{2015}).

\bibitem[{\citenamefont{Pratt et~al.}(2011)\citenamefont{Pratt, Baker,
  Blundell, Lancaster, Ohira-Kawamura, Baines, Shimizu, Kanoda, Watanabe, and
  Saito}}]{Pratt2011}
\bibinfo{author}{\bibfnamefont{F.~L.} \bibnamefont{Pratt}},
  \bibinfo{author}{\bibfnamefont{P.~J.} \bibnamefont{Baker}},
  \bibinfo{author}{\bibfnamefont{S.~J.} \bibnamefont{Blundell}},
  \bibinfo{author}{\bibfnamefont{T.}~\bibnamefont{Lancaster}},
  \bibinfo{author}{\bibfnamefont{S.}~\bibnamefont{Ohira-Kawamura}},
  \bibinfo{author}{\bibfnamefont{C.}~\bibnamefont{Baines}},
  \bibinfo{author}{\bibfnamefont{Y.}~\bibnamefont{Shimizu}},
  \bibinfo{author}{\bibfnamefont{K.}~\bibnamefont{Kanoda}},
  \bibinfo{author}{\bibfnamefont{I.}~\bibnamefont{Watanabe}}, \bibnamefont{and}
  \bibinfo{author}{\bibfnamefont{G.}~\bibnamefont{Saito}},
  \bibinfo{journal}{Nature} \textbf{\bibinfo{volume}{471}},
  \bibinfo{pages}{612} (\bibinfo{year}{2011}), ISSN \bibinfo{issn}{0028-0836}.

\bibitem[{\citenamefont{Han et~al.}(2012)\citenamefont{Han, Helton, Chu,
  Nocera, Rodriguez-Rivera, Broholm, and Lee}}]{Han2012}
\bibinfo{author}{\bibfnamefont{T.-H.} \bibnamefont{Han}},
  \bibinfo{author}{\bibfnamefont{J.~S.} \bibnamefont{Helton}},
  \bibinfo{author}{\bibfnamefont{S.}~\bibnamefont{Chu}},
  \bibinfo{author}{\bibfnamefont{D.~G.} \bibnamefont{Nocera}},
  \bibinfo{author}{\bibfnamefont{J.~A.} \bibnamefont{Rodriguez-Rivera}},
  \bibinfo{author}{\bibfnamefont{C.}~\bibnamefont{Broholm}}, \bibnamefont{and}
  \bibinfo{author}{\bibfnamefont{Y.~S.} \bibnamefont{Lee}},
  \bibinfo{journal}{Nature} \textbf{\bibinfo{volume}{492}},
  \bibinfo{pages}{406} (\bibinfo{year}{2012}), ISSN \bibinfo{issn}{0028-0836}.

\bibitem[{\citenamefont{Balents}(2010)}]{Balents2010}
\bibinfo{author}{\bibfnamefont{L.}~\bibnamefont{Balents}},
  \bibinfo{journal}{Nature} \textbf{\bibinfo{volume}{464}},
  \bibinfo{pages}{199} (\bibinfo{year}{2010}).

\bibitem[{\citenamefont{Yan et~al.}(2011)\citenamefont{Yan, Huse, and
  White}}]{Yan2011}
\bibinfo{author}{\bibfnamefont{S.}~\bibnamefont{Yan}},
  \bibinfo{author}{\bibfnamefont{D.~A.} \bibnamefont{Huse}}, \bibnamefont{and}
  \bibinfo{author}{\bibfnamefont{S.~R.} \bibnamefont{White}},
  \bibinfo{journal}{Science} \textbf{\bibinfo{volume}{332}},
  \bibinfo{pages}{1173} (\bibinfo{year}{2011}).

\bibitem[{\citenamefont{Bramwell and Gingras}(2001)}]{Bramwell16112001}
\bibinfo{author}{\bibfnamefont{S.~T.} \bibnamefont{Bramwell}} \bibnamefont{and}
  \bibinfo{author}{\bibfnamefont{M.~J.~P.} \bibnamefont{Gingras}},
  \bibinfo{journal}{Science} \textbf{\bibinfo{volume}{294}},
  \bibinfo{pages}{1495} (\bibinfo{year}{2001}).

\bibitem[{\citenamefont{Gingras and McClarty}(2014)}]{Gingras_McClarty2013}
\bibinfo{author}{\bibfnamefont{M.~J.~P.} \bibnamefont{Gingras}}
  \bibnamefont{and} \bibinfo{author}{\bibfnamefont{P.~A.}
  \bibnamefont{McClarty}}, \bibinfo{journal}{Reports on Progress in Physics}
  \textbf{\bibinfo{volume}{77}}, \bibinfo{pages}{056501}
  (\bibinfo{year}{2014}).

\bibitem[{\citenamefont{Ross et~al.}(2011)\citenamefont{Ross, Savary, Gaulin,
  and Balents}}]{Ross2011}
\bibinfo{author}{\bibfnamefont{K.}~\bibnamefont{Ross}},
  \bibinfo{author}{\bibfnamefont{L.}~\bibnamefont{Savary}},
  \bibinfo{author}{\bibfnamefont{B.}~\bibnamefont{Gaulin}}, \bibnamefont{and}
  \bibinfo{author}{\bibfnamefont{L.}~\bibnamefont{Balents}},
  \bibinfo{journal}{Phys. Rev. X} \textbf{\bibinfo{volume}{1}},
  \bibinfo{pages}{021002} (\bibinfo{year}{2011}).

\bibitem[{\citenamefont{Sheckelton et~al.}(2012)\citenamefont{Sheckelton,
  Neilson, Soltan, and McQueen}}]{Sheckelton2012}
\bibinfo{author}{\bibfnamefont{J.~P.} \bibnamefont{Sheckelton}},
  \bibinfo{author}{\bibfnamefont{J.~R.} \bibnamefont{Neilson}},
  \bibinfo{author}{\bibfnamefont{D.~G.} \bibnamefont{Soltan}},
  \bibnamefont{and} \bibinfo{author}{\bibfnamefont{T.~M.}
  \bibnamefont{McQueen}}, \bibinfo{journal}{Nat Mater}
  \textbf{\bibinfo{volume}{11}}, \bibinfo{pages}{493} (\bibinfo{year}{2012}),
  ISSN \bibinfo{issn}{1476-1122}.

\bibitem[{\citenamefont{Mourigal et~al.}(2014)\citenamefont{Mourigal, Fuhrman,
  Sheckelton, Wartelle, Rodriguez-Rivera, Abernathy, McQueen, and
  Broholm}}]{Mourigal2014}
\bibinfo{author}{\bibfnamefont{M.}~\bibnamefont{Mourigal}},
  \bibinfo{author}{\bibfnamefont{W.~T.} \bibnamefont{Fuhrman}},
  \bibinfo{author}{\bibfnamefont{J.~P.} \bibnamefont{Sheckelton}},
  \bibinfo{author}{\bibfnamefont{A.}~\bibnamefont{Wartelle}},
  \bibinfo{author}{\bibfnamefont{J.~A.} \bibnamefont{Rodriguez-Rivera}},
  \bibinfo{author}{\bibfnamefont{D.~L.} \bibnamefont{Abernathy}},
  \bibinfo{author}{\bibfnamefont{T.~M.} \bibnamefont{McQueen}},
  \bibnamefont{and} \bibinfo{author}{\bibfnamefont{C.~L.}
  \bibnamefont{Broholm}}, \bibinfo{journal}{Phys. Rev. Lett.}
  \textbf{\bibinfo{volume}{112}}, \bibinfo{pages}{027202}
  (\bibinfo{year}{2014}).

\bibitem[{\citenamefont{Sheckelton et~al.}(2014)\citenamefont{Sheckelton,
  Foronda, Pan, Moir, McDonald, Lancaster, Baker, Armitage, Imai, Blundell
  et~al.}}]{Sheckelton2014}
\bibinfo{author}{\bibfnamefont{J.~P.} \bibnamefont{Sheckelton}},
  \bibinfo{author}{\bibfnamefont{F.~R.} \bibnamefont{Foronda}},
  \bibinfo{author}{\bibfnamefont{L.}~\bibnamefont{Pan}},
  \bibinfo{author}{\bibfnamefont{C.}~\bibnamefont{Moir}},
  \bibinfo{author}{\bibfnamefont{R.~D.} \bibnamefont{McDonald}},
  \bibinfo{author}{\bibfnamefont{T.}~\bibnamefont{Lancaster}},
  \bibinfo{author}{\bibfnamefont{P.~J.} \bibnamefont{Baker}},
  \bibinfo{author}{\bibfnamefont{N.~P.} \bibnamefont{Armitage}},
  \bibinfo{author}{\bibfnamefont{T.}~\bibnamefont{Imai}},
  \bibinfo{author}{\bibfnamefont{S.~J.} \bibnamefont{Blundell}},
  \bibnamefont{et~al.}, \bibinfo{journal}{Phys. Rev. B}
  \textbf{\bibinfo{volume}{89}}, \bibinfo{pages}{064407}
  (\bibinfo{year}{2014}).

\bibitem[{\citenamefont{Haraguchi et~al.}(2015)\citenamefont{Haraguchi,
  Michioka, Imai, Ueda, and Yoshimura}}]{Haraguchi2015}
\bibinfo{author}{\bibfnamefont{Y.}~\bibnamefont{Haraguchi}},
  \bibinfo{author}{\bibfnamefont{C.}~\bibnamefont{Michioka}},
  \bibinfo{author}{\bibfnamefont{M.}~\bibnamefont{Imai}},
  \bibinfo{author}{\bibfnamefont{H.}~\bibnamefont{Ueda}}, \bibnamefont{and}
  \bibinfo{author}{\bibfnamefont{K.}~\bibnamefont{Yoshimura}},
  \bibinfo{journal}{Phys. Rev. B} \textbf{\bibinfo{volume}{92}},
  \bibinfo{pages}{014409} (\bibinfo{year}{2015}).

\bibitem[{\citenamefont{Chen et~al.}(2014)\citenamefont{Chen, Kee, and
  Kim}}]{Chen2014}
\bibinfo{author}{\bibfnamefont{G.}~\bibnamefont{Chen}},
  \bibinfo{author}{\bibfnamefont{H.-Y.} \bibnamefont{Kee}}, \bibnamefont{and}
  \bibinfo{author}{\bibfnamefont{Y.~B.} \bibnamefont{Kim}},
  \bibinfo{journal}{http://arxiv.org/abs/1408.1963}  (\bibinfo{year}{2014}).

\bibitem[{\citenamefont{Chen et~al.}(2016)\citenamefont{Chen, Kee, and
  Kim}}]{Chen2015}
\bibinfo{author}{\bibfnamefont{G.}~\bibnamefont{Chen}},
  \bibinfo{author}{\bibfnamefont{H.-Y.} \bibnamefont{Kee}}, \bibnamefont{and}
  \bibinfo{author}{\bibfnamefont{Y.~B.} \bibnamefont{Kim}},
  \bibinfo{journal}{Phys. Rev. B} \textbf{\bibinfo{volume}{93}},
  \bibinfo{pages}{245134} (\bibinfo{year}{2016}).

\bibitem[{\citenamefont{Anderson}(1987)}]{ANDERSON1987}
\bibinfo{author}{\bibfnamefont{P.~W.} \bibnamefont{Anderson}},
  \bibinfo{journal}{Science} \textbf{\bibinfo{volume}{235}},
  \bibinfo{pages}{1196} (\bibinfo{year}{1987}),
  \eprint{http://www.sciencemag.org/content/235/4793/1196.full.pdf}.

\bibitem[{\citenamefont{Flint and Lee}(2013)}]{Flint2013}
\bibinfo{author}{\bibfnamefont{R.}~\bibnamefont{Flint}} \bibnamefont{and}
  \bibinfo{author}{\bibfnamefont{P.~A.} \bibnamefont{Lee}},
  \bibinfo{journal}{Phys. Rev. Lett.} \textbf{\bibinfo{volume}{111}},
  \bibinfo{pages}{217201} (\bibinfo{year}{2013}).

\bibitem[{\citenamefont{Florens and Georges}(2004)}]{Florens2004}
\bibinfo{author}{\bibfnamefont{S.}~\bibnamefont{Florens}} \bibnamefont{and}
  \bibinfo{author}{\bibfnamefont{A.}~\bibnamefont{Georges}},
  \bibinfo{journal}{Phys. Rev. B} \textbf{\bibinfo{volume}{70}},
  \bibinfo{pages}{035114} (\bibinfo{year}{2004}).

\bibitem[{\citenamefont{Zhao and Paramekanti}(2007)}]{Zhao2007}
\bibinfo{author}{\bibfnamefont{E.}~\bibnamefont{Zhao}} \bibnamefont{and}
  \bibinfo{author}{\bibfnamefont{A.}~\bibnamefont{Paramekanti}},
  \bibinfo{journal}{Phys. Rev. B} \textbf{\bibinfo{volume}{76}},
  \bibinfo{pages}{195101} (\bibinfo{year}{2007}).

\bibitem[{\citenamefont{Fisher et~al.}(1989)\citenamefont{Fisher, Weichman,
  Grinstein, and Fisher}}]{Fisher1989}
\bibinfo{author}{\bibfnamefont{M.~P.~A.} \bibnamefont{Fisher}},
  \bibinfo{author}{\bibfnamefont{P.~B.} \bibnamefont{Weichman}},
  \bibinfo{author}{\bibfnamefont{G.}~\bibnamefont{Grinstein}},
  \bibnamefont{and} \bibinfo{author}{\bibfnamefont{D.~S.}
  \bibnamefont{Fisher}}, \bibinfo{journal}{Phys. Rev. B}
  \textbf{\bibinfo{volume}{40}}, \bibinfo{pages}{546} (\bibinfo{year}{1989}).

\bibitem[{\citenamefont{Herrmann}(1979)}]{Herrmann1979}
\bibinfo{author}{\bibfnamefont{H.}~\bibnamefont{Herrmann}},
  \bibinfo{journal}{Zeitschrift für Physik B Condensed Matter}
  \textbf{\bibinfo{volume}{35}}, \bibinfo{pages}{171} (\bibinfo{year}{1979}),
  ISSN \bibinfo{issn}{0722-3277}.

\bibitem[{\citenamefont{Shao et~al.}(2016)\citenamefont{Shao, Guo, and
  Sandvik}}]{Shao213}
\bibinfo{author}{\bibfnamefont{H.}~\bibnamefont{Shao}},
  \bibinfo{author}{\bibfnamefont{W.}~\bibnamefont{Guo}}, \bibnamefont{and}
  \bibinfo{author}{\bibfnamefont{A.~W.} \bibnamefont{Sandvik}},
  \bibinfo{journal}{Science} \textbf{\bibinfo{volume}{352}},
  \bibinfo{pages}{213} (\bibinfo{year}{2016}).

\bibitem[{\citenamefont{Matsuda and Matsubara}(1957)}]{Matsuda01011957}
\bibinfo{author}{\bibfnamefont{H.}~\bibnamefont{Matsuda}} \bibnamefont{and}
  \bibinfo{author}{\bibfnamefont{T.}~\bibnamefont{Matsubara}},
  \bibinfo{journal}{Progress of Theoretical Physics}
  \textbf{\bibinfo{volume}{17}}, \bibinfo{pages}{19} (\bibinfo{year}{1957}).

\bibitem[{\citenamefont{Santos et~al.}(2004)\citenamefont{Santos, Baranov,
  Cirac, Everts, Fehrmann, and Lewenstein}}]{Santos2004}
\bibinfo{author}{\bibfnamefont{L.}~\bibnamefont{Santos}},
  \bibinfo{author}{\bibfnamefont{M.~A.} \bibnamefont{Baranov}},
  \bibinfo{author}{\bibfnamefont{J.~I.} \bibnamefont{Cirac}},
  \bibinfo{author}{\bibfnamefont{H.-U.} \bibnamefont{Everts}},
  \bibinfo{author}{\bibfnamefont{H.}~\bibnamefont{Fehrmann}}, \bibnamefont{and}
  \bibinfo{author}{\bibfnamefont{M.}~\bibnamefont{Lewenstein}},
  \bibinfo{journal}{Phys. Rev. Lett.} \textbf{\bibinfo{volume}{93}},
  \bibinfo{pages}{030601} (\bibinfo{year}{2004}).

\bibitem[{\citenamefont{Sandvik}(1999)}]{sandvik1999}
\bibinfo{author}{\bibfnamefont{A.~W.} \bibnamefont{Sandvik}},
  \bibinfo{journal}{Phys. Rev. B} \textbf{\bibinfo{volume}{59}},
  \bibinfo{pages}{R14157} (\bibinfo{year}{1999}).

\bibitem[{\citenamefont{Sylju\aa{}sen and Sandvik}(2002)}]{Syljuaasen2002}
\bibinfo{author}{\bibfnamefont{O.~F.} \bibnamefont{Sylju\aa{}sen}}
  \bibnamefont{and} \bibinfo{author}{\bibfnamefont{A.~W.}
  \bibnamefont{Sandvik}}, \bibinfo{journal}{Phys. Rev. E}
  \textbf{\bibinfo{volume}{66}}, \bibinfo{pages}{046701}
  (\bibinfo{year}{2002}).

\bibitem[{\citenamefont{Melko}(2007)}]{Melko2007}
\bibinfo{author}{\bibfnamefont{R.~G.} \bibnamefont{Melko}},
  \bibinfo{journal}{Journal of Physics: Condensed Matter}
  \textbf{\bibinfo{volume}{19}}, \bibinfo{pages}{145203}
  (\bibinfo{year}{2007}).

\bibitem[{\citenamefont{Pollock and Ceperley}(1987)}]{Pollock1987}
\bibinfo{author}{\bibfnamefont{E.~L.} \bibnamefont{Pollock}} \bibnamefont{and}
  \bibinfo{author}{\bibfnamefont{D.~M.} \bibnamefont{Ceperley}},
  \bibinfo{journal}{Phys. Rev. B} \textbf{\bibinfo{volume}{36}},
  \bibinfo{pages}{8343} (\bibinfo{year}{1987}).

\bibitem[{\citenamefont{Dorneich and Troyer}(2001)}]{Dorneich2001}
\bibinfo{author}{\bibfnamefont{A.}~\bibnamefont{Dorneich}} \bibnamefont{and}
  \bibinfo{author}{\bibfnamefont{M.}~\bibnamefont{Troyer}},
  \bibinfo{journal}{Phys. Rev. E} \textbf{\bibinfo{volume}{64}},
  \bibinfo{pages}{066701} (\bibinfo{year}{2001}).

\bibitem[{\citenamefont{Isakov et~al.}(2006)\citenamefont{Isakov, Wessel,
  Melko, Sengupta, and Kim}}]{Isakov2006}
\bibinfo{author}{\bibfnamefont{S.~V.} \bibnamefont{Isakov}},
  \bibinfo{author}{\bibfnamefont{S.}~\bibnamefont{Wessel}},
  \bibinfo{author}{\bibfnamefont{R.~G.} \bibnamefont{Melko}},
  \bibinfo{author}{\bibfnamefont{K.}~\bibnamefont{Sengupta}}, \bibnamefont{and}
  \bibinfo{author}{\bibfnamefont{Y.~B.} \bibnamefont{Kim}},
  \bibinfo{journal}{Phys. Rev. Lett.} \textbf{\bibinfo{volume}{97}},
  \bibinfo{pages}{147202} (\bibinfo{year}{2006}).

\bibitem[{\citenamefont{Nikoli\ifmmode~\acute{c}\else \'{c}\fi{} and
  Senthil}(2005)}]{Nikolic2005}
\bibinfo{author}{\bibfnamefont{P.}~\bibnamefont{Nikoli\ifmmode~\acute{c}\else
  \'{c}\fi{}}} \bibnamefont{and}
  \bibinfo{author}{\bibfnamefont{T.}~\bibnamefont{Senthil}},
  \bibinfo{journal}{Phys. Rev. B} \textbf{\bibinfo{volume}{71}},
  \bibinfo{pages}{024401} (\bibinfo{year}{2005}).

\bibitem[{\citenamefont{Damski et~al.}(2005)\citenamefont{Damski, Fehrmann,
  Everts, Baranov, Santos, and Lewenstein}}]{Damski2005}
\bibinfo{author}{\bibfnamefont{B.}~\bibnamefont{Damski}},
  \bibinfo{author}{\bibfnamefont{H.}~\bibnamefont{Fehrmann}},
  \bibinfo{author}{\bibfnamefont{H.-U.} \bibnamefont{Everts}},
  \bibinfo{author}{\bibfnamefont{M.}~\bibnamefont{Baranov}},
  \bibinfo{author}{\bibfnamefont{L.}~\bibnamefont{Santos}}, \bibnamefont{and}
  \bibinfo{author}{\bibfnamefont{M.}~\bibnamefont{Lewenstein}},
  \bibinfo{journal}{Phys. Rev. A} \textbf{\bibinfo{volume}{72}},
  \bibinfo{pages}{053612} (\bibinfo{year}{2005}).

\bibitem[{\citenamefont{D\"ur et~al.}(2000)\citenamefont{D\"ur, Vidal, and
  Cirac}}]{Durg2000}
\bibinfo{author}{\bibfnamefont{W.}~\bibnamefont{D\"ur}},
  \bibinfo{author}{\bibfnamefont{G.}~\bibnamefont{Vidal}}, \bibnamefont{and}
  \bibinfo{author}{\bibfnamefont{J.~I.} \bibnamefont{Cirac}},
  \bibinfo{journal}{Phys. Rev. A} \textbf{\bibinfo{volume}{62}},
  \bibinfo{pages}{062314} (\bibinfo{year}{2000}).

\bibitem[{\citenamefont{Penrose and Onsager}(1956)}]{Penrose1956}
\bibinfo{author}{\bibfnamefont{O.}~\bibnamefont{Penrose}} \bibnamefont{and}
  \bibinfo{author}{\bibfnamefont{L.}~\bibnamefont{Onsager}},
  \bibinfo{journal}{Phys. Rev.} \textbf{\bibinfo{volume}{104}},
  \bibinfo{pages}{576} (\bibinfo{year}{1956}).

\bibitem[{\citenamefont{Giamarchi et~al.}(2008)\citenamefont{Giamarchi, Ruegg,
  and Tchernyshyov}}]{Giamarchi2008}
\bibinfo{author}{\bibfnamefont{T.}~\bibnamefont{Giamarchi}},
  \bibinfo{author}{\bibfnamefont{C.}~\bibnamefont{Ruegg}}, \bibnamefont{and}
  \bibinfo{author}{\bibfnamefont{O.}~\bibnamefont{Tchernyshyov}},
  \bibinfo{journal}{Nat Phys} \textbf{\bibinfo{volume}{4}},
  \bibinfo{pages}{198} (\bibinfo{year}{2008}).

\bibitem[{\citenamefont{Fisher et~al.}(1973)\citenamefont{Fisher, Barber, and
  Jasnow}}]{Fisher1973}
\bibinfo{author}{\bibfnamefont{M.~E.} \bibnamefont{Fisher}},
  \bibinfo{author}{\bibfnamefont{M.~N.} \bibnamefont{Barber}},
  \bibnamefont{and} \bibinfo{author}{\bibfnamefont{D.}~\bibnamefont{Jasnow}},
  \bibinfo{journal}{Phys. Rev. A} \textbf{\bibinfo{volume}{8}},
  \bibinfo{pages}{1111} (\bibinfo{year}{1973}).

\bibitem[{\citenamefont{Sandvik}(2010)}]{Sandvik2010}
\bibinfo{author}{\bibfnamefont{A.~W.} \bibnamefont{Sandvik}},
  \bibinfo{journal}{AIP Conference Proceedings}
  \textbf{\bibinfo{volume}{1297}}, \bibinfo{pages}{135} (\bibinfo{year}{2010}).

\bibitem[{\citenamefont{Campostrini et~al.}(2006)\citenamefont{Campostrini,
  Hasenbusch, Pelissetto, and Vicari}}]{Campostrini2006}
\bibinfo{author}{\bibfnamefont{M.}~\bibnamefont{Campostrini}},
  \bibinfo{author}{\bibfnamefont{M.}~\bibnamefont{Hasenbusch}},
  \bibinfo{author}{\bibfnamefont{A.}~\bibnamefont{Pelissetto}},
  \bibnamefont{and} \bibinfo{author}{\bibfnamefont{E.}~\bibnamefont{Vicari}},
  \bibinfo{journal}{Phys. Rev. B} \textbf{\bibinfo{volume}{74}},
  \bibinfo{pages}{144506} (\bibinfo{year}{2006}).

\bibitem[{\citenamefont{Carrasquilla and Rigol}(2012)}]{Carrasquilla2012}
\bibinfo{author}{\bibfnamefont{J.}~\bibnamefont{Carrasquilla}}
  \bibnamefont{and} \bibinfo{author}{\bibfnamefont{M.}~\bibnamefont{Rigol}},
  \bibinfo{journal}{Phys. Rev. A} \textbf{\bibinfo{volume}{86}},
  \bibinfo{pages}{043629} (\bibinfo{year}{2012}).

\bibitem[{\citenamefont{Damle and Senthil}(2006)}]{Damle2006}
\bibinfo{author}{\bibfnamefont{K.}~\bibnamefont{Damle}} \bibnamefont{and}
  \bibinfo{author}{\bibfnamefont{T.}~\bibnamefont{Senthil}},
  \bibinfo{journal}{Phys. Rev. Lett.} \textbf{\bibinfo{volume}{97}},
  \bibinfo{pages}{067202} (\bibinfo{year}{2006}).

\bibitem[{\citenamefont{Wu}(1982)}]{Wu1982}
\bibinfo{author}{\bibfnamefont{F.~Y.} \bibnamefont{Wu}}, \bibinfo{journal}{Rev.
  Mod. Phys.} \textbf{\bibinfo{volume}{54}}, \bibinfo{pages}{235}
  (\bibinfo{year}{1982}).

\bibitem[{\citenamefont{Bonnes and Wessel}(2011)}]{Bonnes2011}
\bibinfo{author}{\bibfnamefont{L.}~\bibnamefont{Bonnes}} \bibnamefont{and}
  \bibinfo{author}{\bibfnamefont{S.}~\bibnamefont{Wessel}},
  \bibinfo{journal}{Phys. Rev. B} \textbf{\bibinfo{volume}{84}},
  \bibinfo{pages}{054510} (\bibinfo{year}{2011}).

\bibitem[{\citenamefont{Jo et~al.}(2012)\citenamefont{Jo, Guzman, Thomas,
  Hosur, Vishwanath, and Stamper-Kurn}}]{Gyu-Boong2012}
\bibinfo{author}{\bibfnamefont{G.-B.} \bibnamefont{Jo}},
  \bibinfo{author}{\bibfnamefont{J.}~\bibnamefont{Guzman}},
  \bibinfo{author}{\bibfnamefont{C.~K.} \bibnamefont{Thomas}},
  \bibinfo{author}{\bibfnamefont{P.}~\bibnamefont{Hosur}},
  \bibinfo{author}{\bibfnamefont{A.}~\bibnamefont{Vishwanath}},
  \bibnamefont{and} \bibinfo{author}{\bibfnamefont{D.~M.}
  \bibnamefont{Stamper-Kurn}}, \bibinfo{journal}{Phys. Rev. Lett.}
  \textbf{\bibinfo{volume}{108}}, \bibinfo{pages}{045305}
  (\bibinfo{year}{2012}).

\end{thebibliography}

\end{document}